# VIS spectroscopy of NaCl – water ice mixtures irradiated with 1 and 5 keV electrons under Europa's conditions: Formation of colour centres and Na colloids

R. Cerubini[1], A. Pommerol[1], A. Galli[1], B. Jost[2], P. Wurz[1], N. Thomas[1]

1) Physikalisches Institut, University of Bern, Switzerland

2) Jet Propulsion Laboratory, California Institute of Technology, 4800 Oak Grove Drive, Pasadena, CA 91109, USA

*) Corresponding author, (romain.cerubini@unibe.ch)

## Abstract

Recent laboratory efforts and telescopic observations of Europa have shown the relevance of a yellow colouration of sodium chloride (NaCl) caused by crystal defects generated by irradiation. We further investigate this process by irradiating (with energetic electrons) different types of analogues where NaCl is associated in different ways to water ice. We produce two types of icy analogues: compact slabs and granular particles where we investigate two particle sizes (5 and 70 µm). We perform electron irradiation at cryogenic temperatures (100 K) and under high vacuum ($10^{-7}$ mbar) conditions, with energies of 1 and 5 keV. We observe the formation of two different types of colour centres. The so-called F-centres (460 nm) were formed in every sample, but the intensity of the absorption band within the compact slabs surpassed any other icy analogues and was comparable to the intensity of the absorption band within pure NaCl grains. M-centres (720 nm) have not been detected at the surface of Europa so far, and were close to the detection limit during our irradiation of compact slabs. The slabs could be good analogues for Europa's surface as they produce mainly F-centres. Other notable differences have been observed between compact slabs and granular samples, such as the presence of an absorption band at 580 nm attributed to colloids of Na, exclusively within granular samples. Such absorptions have not been reported in previous studies.

# 1. Introduction

The icy moons of the outer Solar System have already been studied for several decades, starting with the Pioneer and Voyager missions in the 1970s, which provided the first close-up views of their surfaces. Voyager 2 performed a close flyby of Europa in July 1979, which revealed its complex surface (Johnson et al., 1983; Smith et al., 1979). Between 1996 and 2003, measurements of the gravitational parameters and the magnetic properties of Europa by NASA's Galileo mission (Anderson et al., 1998; Khurana et al., 1998; Kivelson, 2000; Schubert et al., 2004) provided the first constraints on the interior composition and indicated the presence of a global ocean beneath the global icy shell. Surface geomorphologies have also been suggested as evidence for the presence of subsurface water (Carr et al., 1998; Head and Pappalardo, 1999; Pappalardo et al., 1999). The presence and stability of liquid water in the solar system are of scientific interest because of their implications for habitability (Chyba and Hand, 2005; Chyba and Phillips, 2002, 2001) and the key role water plays in planetary formation and evolution due to its unique physical and chemical properties.

The induced magnetic field requires a conductive global ocean, and geochemical modelling suggests that the ocean contains alkali ions, sulphur, chloride and magnesium compounds in various proportions (Zolotov and Kargel, 2009; Zolotov and Shock, 2001). Analysis of the exosphere of the icy moon tends to confirm the endogenic source of alkali ions (Brown, 2001; Ozgurel et al., 2018). In recent studies, the potential evolution of Europa's ocean was modelled as a function of pH, temperature and composition (Johnson et al., 2019; Vu et al., 2016), reinforcing the case for the presence of Mg, Na, Cl and sulphates from endogenic sources as ions within the salty ocean.

Since the late 1990s, numerous experimental studies have been conducted to constrain the composition of the non-icy components at the surface of Europa. Carlson and co-authors (1999, 2002, 2009) showed, through laboratory and modelling, that the presence of hydrates of sulphuric acid could explain features in the near-infrared spectrum of the trailing hemisphere of Europa. The presence of Mg-bearing species has also been investigated in detail. The first laboratory results with brines of magnesium and sodium sulphate were reported by (Dalton, 2007, 2003; Dalton et al., 2005; McCord et al., 1999, 1998, 2002). More recently, linear mixing reflectance models supported by results from laboratory experiments have indicated the presence of chlorides (Hanley et al., 2014, 2012). (Ligier et al., 2016) reported new observations of Europa's surface and concluded that chlorides could spectroscopically match NIR features on the trailing side. (Trumbo et al., 2019) identified an absorption feature around 450 nm in the reflectance spectra of chaotic terrains on the leading side of Europa, which is well-reproduced in laboratory spectra of NaCl grains irradiated by energetic particles.

Several geological mechanisms are thought to bring water from the subsurface ocean to the surface. These include tectonism, diapirism, surface effusion, cryo-volcanism and plumes (Fagents, 2003; Howell and Pappalardo, 2018; Prockter and Patterson, 2009). Observations from the Cassini spacecraft have demonstrated the presence of plumes at Enceladus (Nimmo et al., 2007; Postberg et al., 2009) as sources of pristine surface material from a liquid ocean, or from sub-surface liquid reservoirs closer to the surface. Recent observations of Europa (Jia et al., 2018; Roth et al., 2014; Sparks et al., 2016) also suggest the presence of plumes or geysers.

As soon as fresh particles are emplaced at the surface of a Jovian moon, they become subject to the effects of the intense magnetosphere of Jupiter. In particular, a constant bombardment by an enormous flux of energetic particles with a broad energy distribution occurs. Ion and electron bombardment can lead to radiolysis, to implantation of ions and material from meteorites impacts mixed with native surface material (Alvarellos et al., 2008; Johnson et al., 2004; Paranicas et al., 2009, 2001; Zahnle et al., 2008, 2003) and to potential chemical reactions arising from radiolytic processes (Cooper et al., 2001; Shirley et al., 2010). These processes can be simulated in the laboratory to better understand the effects of the interaction between pure water ice and the incoming flux of energetic particles in terms of sputtering and disruption of the surface (Cassidy et al., 2013; Galli et al., 2018b, 2018a, 2017; Teolis et al., 2005). The interactions generate $H_2$, $O_2$, $H_2O_2$, $H_2O$ and other minor components can produce exospheres enriched in these species (Plainaki et al., 2020, 2010).

(Hand and Carlson, 2015) have recently proposed NaCl as a probable candidate for the non-icy surface material because of the radiation-induced colouration of the crystal by electron bombardment (Bridges et al., 1990; Compton and Rabin, 1964; Hartog et al., 1994; Schwartz et al., 2008; Seitz, 1946), which resembles the yellowish tones observed over a large fraction of Europa’s surface. The formation of crystal defects upon irradiation leads to the presence of the so-called F- and M-centres, which produce visible colouration as a result of absorption features at 460 and 720 nm, respectively. F- and M-centres are generated by the replacement of a $Cl^-$ ion by, respectively, one or two electrons in a vacant $Cl^-$ site within the crystal. (Poston et al., 2017) performed a more detailed study based on the results from (Hand and Carlson, 2015) on pure NaCl grains. Neither of these two studies have shown the presence of Na colloids at temperatures relevant to the surface of Europa (around 100K). This particular defect, generating an absorption feature at 580 nm, is the result of the agglomeration of Na atoms within the crystal. It is a consequence of the formation of the F-centre and of the mobility of crystalline defects within the lattice (Sugonyako, 2007).

Frozen brines have been proposed in previous studies as relevant analogues for the surface, placing ions, hydrates and pure salts in a water-ice matrix (Carlson et al., 2005; McCord et al., 2002; Orlando et al., 2005). Flash-freezing of solutions was used in these experimental studies to produce macroscopic analogue samples; mimicking the flash-freezing process expected to happen at the surface, following brine mobilisation within geysers. Nevertheless, some of the pristine material at the surface of Europa could also result from slow crystallisation processes occurring in the sub-surface or in a hydrogeological-like system in the ice shell before being exposed to the surface by tectonic processes.

While electron irradiation of pure NaCl samples shows convincing colouration which could explain some of the visible spectral features seen at the surface of Europa, this process has not yet been tested with realistic analogue materials, combining NaCl and water ice. Here, we report on the results of irradiation experiments performed with different types of associations between water ice and sodium chloride under pressure and temperature conditions relevant for Europa. The behaviour of granular ice samples produced by flash-freezing is systematically compared to compact slabs produced by slower crystallisation of the ice and the salts. The interpretations of the resulting visible spectra are discussed in terms of their implications for the analysis of existing data and future missions (ESA's JUICE and NASA's Europa Clipper).

# 2. Methods

## 2.1. Samples production

When producing salty ices as analogues, our main concerns are the representativeness of our samples for different plausible processes at the surface of Europa, and the good characterisation and reproducibility of the samples. We produce salty ice particles with two setups developed in the Planetary Ice Laboratory of the University of Bern, the Setup for Production of Icy Planetary Analogues (SPIPA) version –A and –B (Jost et al., 2016; Poch et al., 2016a; Pommerol et al., 2019; Yoldi et al., 2015). These devices have already been used for the production of analogues for different types of Solar System objects such as comets, by mixing ice with black carbon and mineral dust (Jost et al., 2017b, 2017a, 2016; Pommerol et al., 2015) as well as complex organics (Poch et al., 2016a, 2016b). It has also been used for Mars-related studies to assess the influence of the mixing mode between ice and dust (Yoldi et al., 2021), and for the establishment of a dielectric model for subsurface sounding of the Martian surface (Brouet et al., 2018). Jost et al. (2016) have also noted the resemblance between the reflectance phase curves of these particles and the observed properties of the surfaces of Europa and Enceladus. The SPIPA setups have therefore been selected for a series of studies of ice sputtering in the context of icy moons surfaces (Galli et al., 2018b, 2017, 2016). They were also used for the polarimetric study of icy particles to provide insight into the physical properties of icy surfaces (Poch et al., 2018). In addition to the reproducibility and good characterisation of the SPIPA products, their ability to produce ice particles with variable amounts of insoluble and soluble contaminants make them a logical choice to produce the analogues required for this study.

The SPIPA-A setup produces spherical particles of water ice with a grain size distribution of 4.5 ± 2.5 µm. The principle is described in Figure 2 of (Pommerol et al., 2019). Small liquid droplets generated by an ultrasonic nebuliser freeze in cold air (~200K) and sediment into an aluminium bowl cooled by liquid nitrogen.

The SPIPA-B setup produces spherical particles of water ice with a size distribution of 70 ± 30 µm. As in SPIPA-A, liquid droplets are produced by an ultrasonic nebuliser, but in this setup, they are directed into a bowl filled by liquid nitrogen and freeze at its surface before sinking into it.

For both SPIPA-A and SPIPA-B, different saline solutions can be pumped into the setup and frozen for further analysis and experiments. We have produced salty SPIPA-A and SPIPA-B ices from aqueous solutions of NaCl in various concentrations to assess the effect of the amount of salt within a particle, and the influence of grain size on their colouration when subjected to electron irradiation. SPIPA-B can produce ice particles from a broad range of concentrations including saturated solutions under standard laboratory conditions (over 26 wt% for NaCl). Therefore, concentrations of 5, 10 and 30 wt% were produced. The SPIPA-A setup is more restricted with regard to the maximum amount of dissolved salt, and only 5 and 10 wt% solutions have been used to produce ice particles.

We refer to samples prepared from solutions of NaCl as "intra-mixtures" in which salt crystals are embedded into a matrix of ice within each particle. We also produced "inter-mixtures" to vary the texture and mixing mode of our samples and to study their influence on the outcome of electron irradiation. These samples consist of intimate mixtures of pure grains of SPIPA-B ice and NaCl (here sieved to < 100 µm) prepared following the procedures described in (Jost et al., 2017b, 2017a; Poch et al., 2016a; Yoldi et al., 2015), using aluminium bottles pre-cooled with liquid nitrogen containing, separately, the water ice and the salt. Both are hereafter mixed within a third bottle thank to a vibrating vortex ("inter-SPIPA B 10 and 30 wt%" in Figure 4, A).

The third type of sample texture investigated consists of compact slabs prepared by slowly cooling down solutions of distilled water and NaCl in various concentrations. We prepared solutions at ambient conditions, then placed them within aluminium sample holders in a freezer at 235 K to let them crystallise slowly (20 K/hrs; 0.005 to 0.01 K/sec) following thermodynamic equilibrium.

## 2.2. MEFISTO Chamber

The MEFISTO (MEsskammer für FlugzeitInStrumente und Time-Of-Flight) facility at the University of Bern was developed originally to test and calibrate mass spectrometers for various space missions (Galli et al., 2016; Hohl, 2002; Marti et al., 2001). It consists of a vacuum chamber with 1.6 $m^3$ volume, able to reach $10^{-8}$ mbar of pressure with an ice sample, coupled to an ion cyclotron source. The chamber is equipped with a table fixed onto a hexapod mechanism to move it precisely in the *x*, *y* and *z* directions (Figure 1). The chamber has recently been equipped with an electron gun to irradiate samples with electrons of energies ranging from 100 eV to 10 keV.

In this study, we have used two energies: 1 and 5 keV to see the effect of accelerated electrons on crystal defects formation from a spectroscopic point of view. These energies are relevant considering the important range of energies (from 100 eV to MeV) associated with particles that hit the surfaces of Galilean moons within the Jupiter magnetosphere (Cooper et al., 2001; Paranicas et al., 2009, 2002, 2001). We have chosen currents of 1 and 3 µA to reach sufficient doses to produce surface reactions ($10^{16}$ $e^-/cm^2$) and to mimic longer irradiation period at the surface of Europa while avoiding or minimising the sublimation of water ice at the irradiated location (Galli et al., 2018b). The irradiation conducted at 3 µA can be compared to 3 times longer irradiation at 1 µA. We have irradiated for 15 minutes on each sample and in each configuration, reaching total doses between $10^{16}$ and $10^{17}$ $e^-/cm^2$ (or fluxes of $10^{13}$ to $10^{14}$ $e^-/cm^2.s$). Compared to the values of electron fluxes reaching the surface of Europa of $3 \cdot 10^8$ $e^-/cm^2.s$ (Cooper et al., 2001; Paranicas et al., 2009), our electron fluxes correspond to 1 to 10 years at the surface. While our total doses are representative of the short-term evolution of the surface of Europa, our fluxes are too high by a factor of about $10^6$. We acknowledge that this leads to possible caveats in the interpretation of the experimental results in this work and in past studies performed with similar fluxes. Indeed, threshold effects and competitions between processes with very different kinetics might make the comparison to icy moons surface problematic. Using high fluxes in experiments is however unavoidable as realistic fluxes would require very long experiments that are hardly feasible technically and would consume enormous resources. Physical models of the processes at play are probably required in the future to better extrapolate the results from the lab to icy moons surfaces.

The two parameters strongly affecting the electron flux are the beam current, that we can adjust, and the beam diameter, which is affected by the desired energy of the electrons. The penetration depth of electrons in matter varies with electron energy. It has been reported by (Hand and Carlson, 2011; Johnson, 1990) for 1 keV, being 46 nm and for 5 keV, being 550 nm with a 45° of angle between the electron source and the surface of pure water ice samples.

The beam current was measured before and after each irradiation run to establish the electron flux ($e^-/cm^2$) with a relative precision of 22%. The error is coming from variabilities of the setup in the beam current as well as uncertainties on the beam diameter. The latter one was established by measuring the intensity of the beam on a copper plate covered with black aluminium tape (Figure 1, B). The copper plate act as a faraday cup to protect the sample of ejected electrons (secondary electrons) from the irradiated surfaces (metals or icy samples) (Jurac et al., 1995), The number of secondary electrons is negligible compared to the number of electrons generated by the electron gun (Galli et al., 2018b). Knowing the dimension of the plate and the maximum intensity of the beam, and performing incremental displacements while continuously measuring the intensity, we could determine the beam diameter within a 10% relative uncertainty. Increasing the energy of the electrons from 1 keV to 5 keV results in a decrease of the beam size. It happened only once that the electron flux at 1 and 3 µA were similar (around $6{,}0.10^{16}$ $e^-/cm^2$ total deposition; slabs showed as green squares and red symbols, Figure 11), because of a decreasing current during the irradiation (from the initial 3 µA) .

The sample holder (an aluminium cylinder of 4 cm diameter and 0.9 cm deep, Figure 1 and 2) is firmly fixed onto the table cooled by a circuit of liquid nitrogen to maintain its temperature around 100K with an expected gradient of temperature of 10 to 20 K at maximum between the bottom and the top of the sample holder. Two pressurised tanks of liquid nitrogen outside the chamber are used to cool the table. We have used two configurations for the cooling:

1. The continuous flow of nitrogen between the two tanks and through the cooling circuit in the table to ensure the cooling of the sample. The circulation of liquid nitrogen is maintained by establishing a differential pressure between the tanks. The direction of circulation is changed each time a tank is fully filled with the nitrogen. This is the configuration for which the temperature is the lowest and the most constant, but it requires frequent manual interventions. It is used for most of the irradiation experiments. During the irradiation experiments, the temperature of the sample holder is monitored continuously, indicating a sample holder temperature of -180 ± 2°C (93 K).
2. Automatic re-filling of the cooling circuit from a single pressurised tank as soon as the measured temperature of the sample holder increases. This cooling mode is fully automatic but results in periodic variations of the temperature. It is therefore used to preserve the sample overnight between experiments and measurements campaigns.

## 2.3. MoHIS system

All the optical characterisations of the samples within the simulation chamber were performed with the Mobile Hyperspectral Imaging System (MoHIS), which is installed on one of the side flanges of the chamber (Figure 1). This hyperspectral imaging system derives directly from the one previously in use on our SCITEAS simulation chamber (Jost et al., 2017b; Poch et al., 2016b, 2016a; Pommerol et al., 2015; Yoldi et al., 2021) but has been rebuilt and reconfigured to be mobile and used with different simulation chambers or as a standalone system on a table. MoHIS consists of a monochromatic light source and two cameras. The light is generated by a QTH Source (50 – 250 W, F2/2 FSA, Newport©) and focused with a parabolic mirror into an Oriel MS257 grating monochromator (Newport©). A narrow wavelength band (FWHM of either 6.5 or 13 nm) is isolated within the VIS-NIR (350 - 2500 nm) spectral range using a selection of refractive gratings and low-pass transmission filters. The monochromatic light is then conducted via a large optical fibre bundle (Bundle 500, CeramOptec©) and illuminates the sample through a quartz window. A fused-silica lens is used to adjust the divergence of the illuminating beam and hence the size of the illuminated spot on the samples. Two cameras (a visible CCD camera, model 1501M from Thorlabs and a near-infrared Mercury-Cadmium-Telluride camera, model Xeva2.5 from Xenics) are fixed into a light and stable aluminium structure and look at the sample inside the chamber through a large quartz window and via two folding mirrors. This structure also holds the tip of the fibre bundle and the focusing lens and is directly fixed onto a standard DN160CF flange of the chamber equipped with a quartz window. The rest of the equipment (light source, electronics, computer) is installed onto a trolley and connected to the box on the flange by optical fibre bundle and cables.

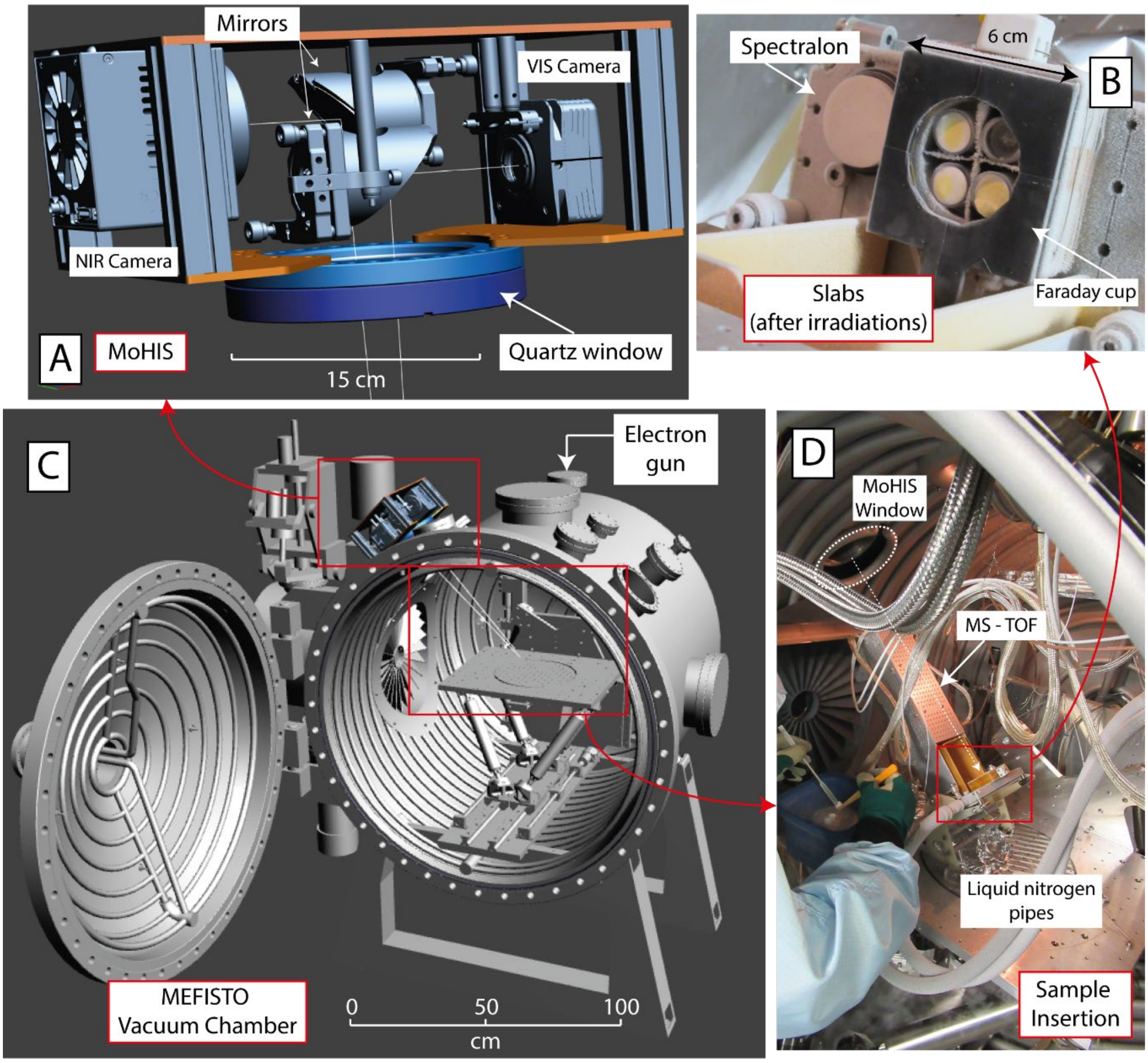

*Figure 1: A) CAD of the MoHIS cameras positioned on the side flange of the MEFISTO vacuum chamber. B) close-up view of the samples (here irradiated slabs) in position on the cooled-down table, after re-opening the chamber. C) CAD of the MEFISTO vacuum chamber with the position of the electron gun and the MoHIS cameras. D) Picture of the inside of the MEFISTO chamber during sample insertion. The mass-spectrometer – Time of flight (MS-TOF) is not described in this paper (will be the subject of another publication).*

Because of the construction of the chamber and locations of the flanges, the sample must be moved between two different positions to be either irradiated with electrons or imaged by MoHIS in optimal illumination conditions. Controlled and repeatable movements of the sample are achieved using a high-precision hexapod table. The error positioning of the hexapod is in the order of micrometres.

The measurement sequence consists of repeatedly shifting the wavelength transmitted by the monochromator and subsequently acquiring an image with one of the two cameras, over a chosen spectral range and with a chosen spectral sampling. Following a number of improvements compared to our previous work, we can now measure an entire sequence (0.4 to 2.5 µm with a spectral sampling of 15 nm in the VIS and 6 nm in the NIR) in less than 30 minutes. Hyperspectral cubes are then constructed by superposing the images taken at all different wavelengths. *Regions Of Interest* (ROIs) can be defined manually during the calibration process, to extract an arbitrary region of the sample and calculate the average and standard deviation of all spectra within this region.

The values of reflectance are calibrated in units of reflectance factor **REFF** (Hapke, 2012) which is the ratio between the radiance scattered by the sample in a given geometry to the radiance scattered by a perfectly diffusive Lambertian surface in the same geometry. It is defined as follows:

$$REFF = \pi \frac{I}{F \cos(i)}$$

Where $I$ is the radiance in [$W \cdot m^{-2} \cdot sr^{-1}$] (the scattered light from the sample), $F$ the irradiance in [$W \cdot m^{2}$] (the incoming light) and $i$ the incidence angle of the incoming light.

To calibrate our measurements, we systematically perform before and/or after the measurements of a sample, a sequence of reference measurements with a plate of Spectralon™ (Labsphere). This material shows a hemispheric reflectance of 0.99 over the VIS-NIR spectral range with a nearly-Lambertian behaviour, in particular at low phase angle. A shallow absorption at 2.2 µm is corrected using manufacturer-provided data. The Spectralon plate is placed at the exact same location as the sample holder, and hyperspectral cubes are acquired with the exact same configuration of illumination and imaging as for the sample. The final calibration consists of a division of the hyperspectral cube obtained with the sample by the hyperspectral cube obtained with the spectralon plate.

The acquisition of hyperspectral cubes for both the sample and the Spectralon also include the measurement of a number of "dark images" at regular intervals. These images are obtained by closing a physical shutter within the monochromator before acquiring the image of the unilluminated scene. These dark images are then interpolated and subtracted from the data to correct for unwanted variations of the dark signal in the cameras during the acquisition of the cubes. This is particularly useful with the data of the cooled mercury-cadmium-telluride (MCT) infrared camera, which shows high sensitivity to temperature variations in the laboratory.

Repeated measurements realised with fresh and stable samples reveal a maximum relative variability of 3% in the VIS over more than 10 hours. This variability is the result of two main sources, the sample and the setup. If the sample shows macroscopic roughness or small-scale heterogeneities, the definition of different ROIs will produce some statistical variability. When unaltered and measured under stable conditions, our samples are, however, rather homogeneous. Errors from the setup can originate from the monochromator (reproducibility of the gratings positioning), from the lamp (stability of the light flux) or from the camera (dark current, averaging temporal variability, altered pixels on CCD).

Band depths ($1-R_{band}/R_{continuum}$) are computed after normalisation of all spectra at 920 nm. The effects of the electron irradiation are highlighted by computing spectral ratios between the spectra of irradiated samples and the spectra of pristine samples. The latter either come from a different ROI within the same cube (if the size of the sample allowed selecting a region not affected by the electron irradiation) or from the cubes measured prior to any irradiation. We have systematically tested both cases on 85% of the dataset and found a variability from the error propagation contained within 2% (with one outlier value at 5%, corresponding to a slab experiment where band depths are considerably larger than in other samples). This verification shows that the irradiation produces local effects while non-irradiated areas of the sample nearby remain unaltered in terms of colour centres.

The error bars given for the band depths computation represent the variability of the intensity of absorption features within the manually defined ROIs (i.e. the variability of the sample itself due to irradiation) and not the accuracy of the reflectance values. The normalisation at 920 nm allows us to standardise each measurement before computing band depth. This chosen wavelength is a compromise between a good signal-to-noise ratio (which decreases toward longer wavelengths) and the distance in the spectrum from the absorptions we study at shorter wavelengths.

When selecting the ROIs strictly within the irradiated area, the 1σ error is estimated to be 5%, based on the standard deviation within the different pixels of the ROI compared to the average signal within the ROI. This error is encompassing several sources, including the intrinsic variability of the signal within the ROI, therefore linked to the heterogeneity of the sample. Significant care has been taken to select ROIs within the irradiated area.

Some calibration artefacts are observed and are sometimes difficult to remove (e.g. small bump present at 695 nm (Figure 6, *f* and *g*)). They are caused by the grating switch of the monochromator during the hyperspectral cube acquisition, which is not completely reproducible. For the band depths computation, we select wavelengths to avoid this artefact and prevent any bias. An absorption band is nevertheless present at 715 nm as discussed in section 3 and 4. This particular feature is not produced by calibration issues; the correction that can be applied at 700 nm to correct the displacement induced by the grating switch is in the order of 3% and would just slightly flatten it.

As explained in section 2.2, the impacting electrons only affect the first ten to hundreds of nanometres at the surface of the ice, which is less than the thickness on which the scattering of the visible light occurs (up to a hundred particle diameters and therefore several hundreds of micrometres up to a few millimetres, Hapke, 2012). Therefore, in addition to the areal variability of the irradiation, a thickness variability is also present, and the irradiated samples should be considered as layered with an irradiated top layer superposed onto less altered layers.

Table 1 summarises the irradiation parameters used during the experiments.

*Table 1: Main parameters used for the electron irradiations. SPIPA-B inter, SPIPA-B and SPIPA-A are all granular samples. Every sample has been irradiated at 1 and 5 keV.*

| | Sample type | [NaCl] (wt%) | Energies and Currents | Beam diameter (cm) | Irradiation Time (sec) | | Averaged electron fluxes ($e^-/cm^2$) |
|---|---|---|---|---|---|---|---|
| Granular samples | **Inter mixtures SPIPA-B (67 µm)** | 10 wt%, 30 wt% | 1 keV 1 µA, 5 keV 1 µA | 1 keV: 0,5 to 0,7 | 900 | Granular samples | *1 keV - 1 µA*: 2,09·$10^{16}$ |
| | **Intra mixtures SPIPA-B (67 µm)** | 30 wt% | 1 keV 1 µA, 5 keV 1 µA | | | | *5 keV - 1 µA*: 2,95·$10^{16}$ |
| | | 10 wt%, 5wt% | 1 keV 1 µA, 5 keV 1 µA, 1 keV 3 µA, 5 keV 3 µA | | | | *1 keV - 3 µA*: 5,38·$10^{16}$ |
| | **Intra mixtures SPIPA-A (5 µm)** | 5 wt%, 10 wt% | 1 keV 1 µA, 5 keV 1 µA, 1 keV 3 µA, 5 keV 3 µA | | | | *5 keV - 3 µA*: 1,09·$10^{17}$ |
| | **Slab** | 5 wt% | 1 keV 1 µA, 5 keV 1 µA | 5 keV : 0,3 to 0,6 | | Slabs | *1 keV - 1 µA*: 2,81·$10^{16}$; *5 keV - 1 µA*: 5,94·$10^{16}$ |
| | | 10 wt%, 20 wt%, 30 wt% | 1 keV 1 µA, 5 keV 1 µA, 1 keV 3 µA, 5 keV 3 µA | | | | *1 keV - 3 µA*: 5,84·$10^{16}$; *5 keV - 3 µA*: 1,49·$10^{17}$ |

# 3. Results

## 3.1. Overview

The reflectance spectra of the inter-mixtures of SPIPA-A and -B samples (group A, granular ices: inter-mixtures, pictures in Figure 2) are shown in Figure 4. The reflectance spectra of the intra-mixtures of SPIPA-A and -B samples (group A and B) are shown respectively in the Figure 5 and Figure 6. The reflectance spectra of the slabs are shown in the Figure 7 . The first irradiations were always performed on pristine ice. Due to the small size of the samples, the second irradiation consisted of a re-irradiation of the previous spot.

The RGB colour composites presented in Figure 2 show a pronounced yellow colouration on the slab samples and, to a lower extent at the surface of granular ices.

The spectra shown in this section are averaged over ROIs manually defined to surround the areas most affected by irradiation. Obviously, these average reflectance spectra encompass an intrinsic spatial variability that reflects the complexity and heterogeneity of the irradiated samples.

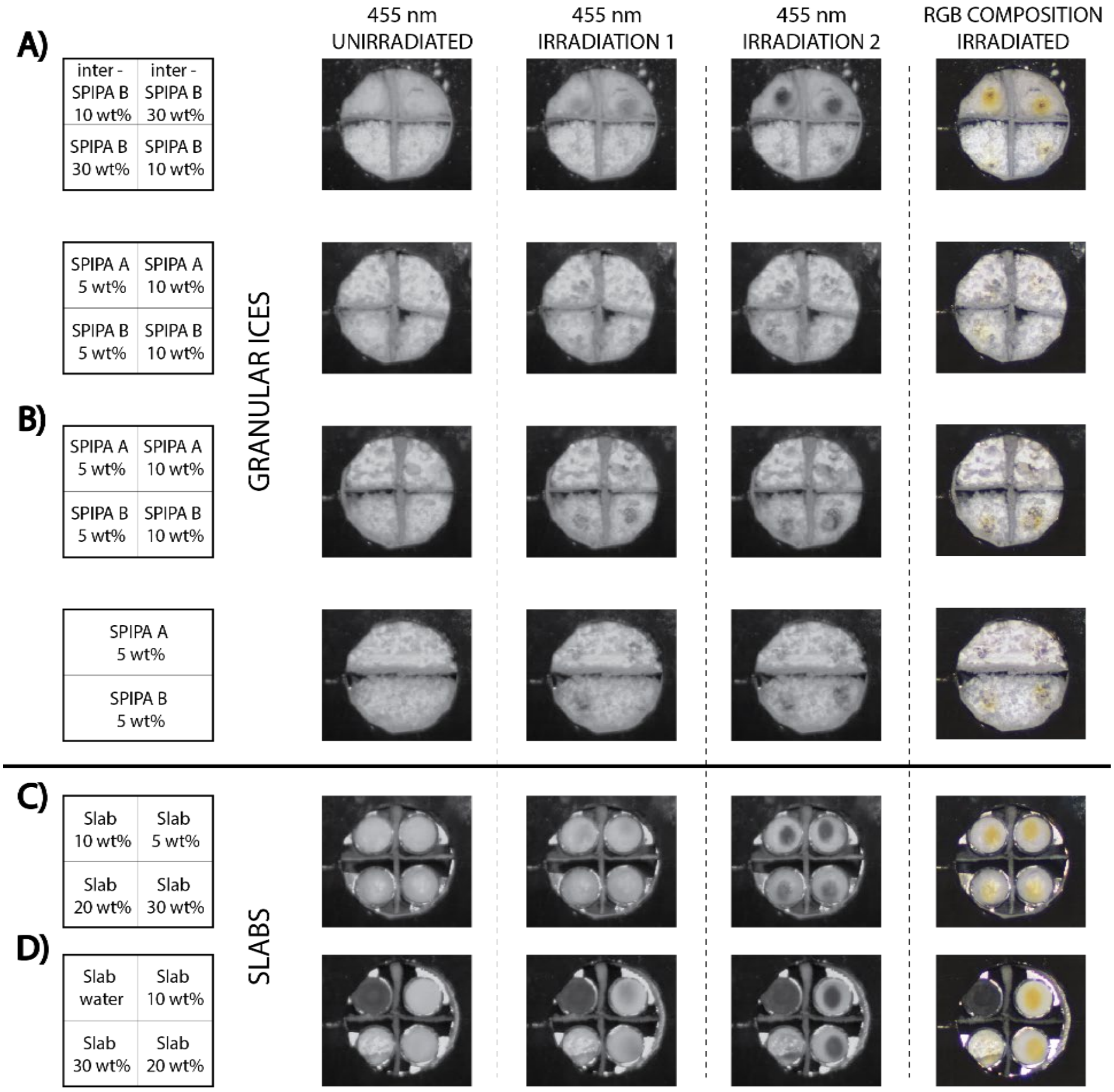

*Figure 2: The entire set of samples irradiated in this study. Granular ices are on the top panels and compact slabs on the bottom panel. From left to right, the columns correspond to the composition of the sample, then three columns with pictures taken at 455 nm prior to any irradiation, after first irradiation, and after second irradiation, respectively. The last column shows RGB colour composites (B: 395-500 nm, G: 500-605 nm, R: 605-695 nm) of the samples after irradiation. One can notice the yellow colouration induced by the electron irradiation as well as the stronger effect on the slabs compared to the granular ices.*

## 3.2. Colour centres formation in crystal lattice

Irradiation of alkali halide crystals generates crystal defects known as colour centres. Figure 3 shows the defect induced by electron irradiation within the crystal lattice of a NaCl crystal. Details of the processes to form such defect are discussed in section 4. The scheme is here to help the reader to link absorption features and crystal defects.

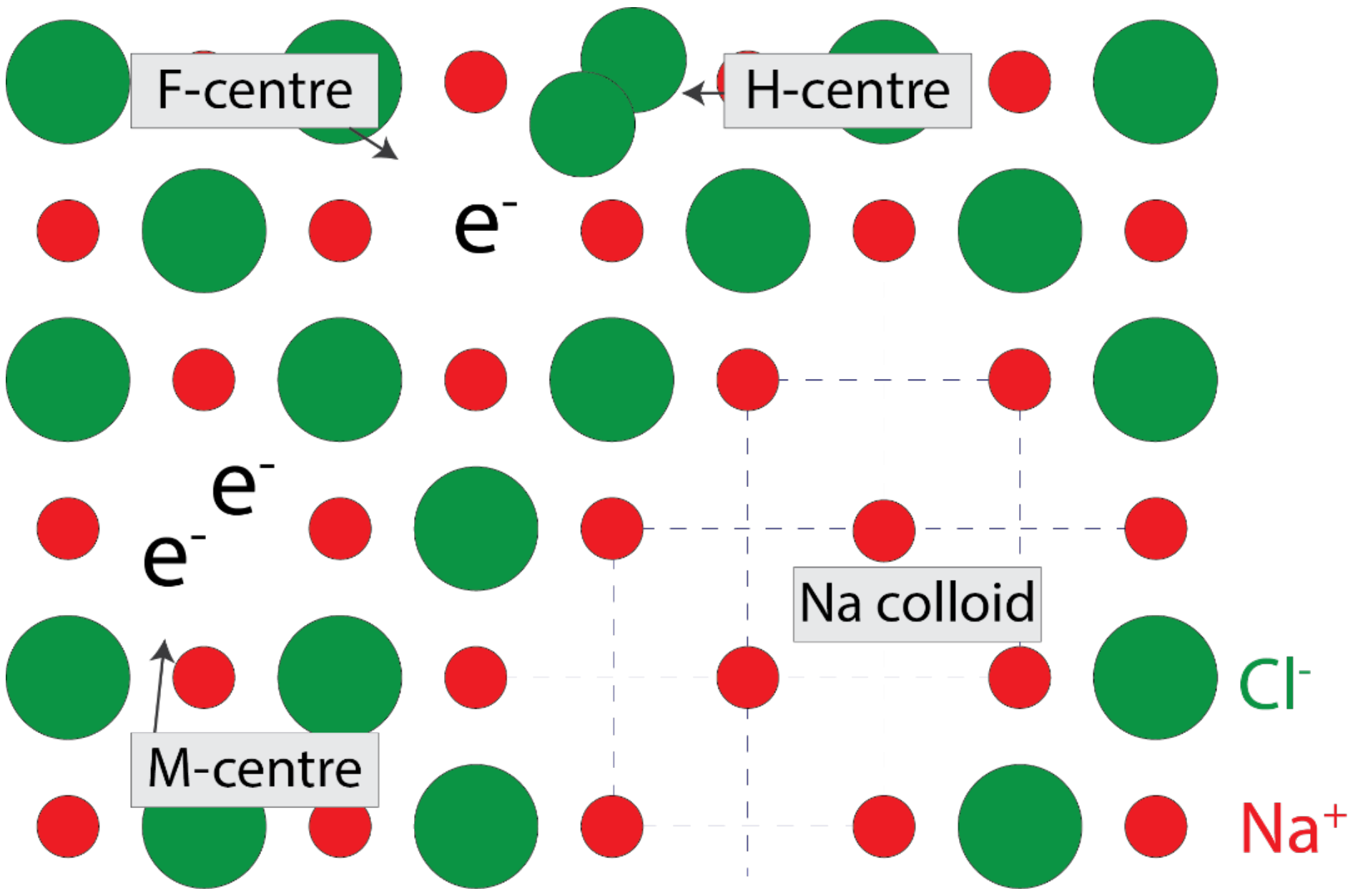

*Figure 3: Scheme of a NaCl crystal lattice affected by electron irradiation. Chloride anions are indicated in green and sodium cations are in red. The F- and M- colour centres are both generating absorption in the VIS (460 nm and 720 nm) as well as the Na colloid (580 nm). The H-centre does not produce any absorption in the VIS. Inspired from* (Hartog et al., 1994; Soppe et al., 1994; Sugonyako, 2007).

For the scope of this study, the absorption we are interested in are located at 460, 580 and 720 nm. With our experimental settings, we measure exactly at 455, 575 and 710 nm. With our FWHM, these bands correspond to absorption encompassing 448,5 to 461,5 nm for the 460-nm band, 568,5 to 581,5 nm for the 580-nm band and 704, 5 to 716, 5 nm for the 720-nm band.

## 3.3. Irradiation of granular ices

### 3.3.1. Inter-mixtures

We have irradiated SPIPA-B inter-mixture samples with 1 µA at 1 and 5 keV electrons. The results presented in Figure 4 show the formation of F-centre defects (absorption feature at 460 nm) from the first irradiation onwards. The intensity of the absorption feature varies only slightly between the two concentrations, being a little stronger for the 30wt% case.

The secondary irradiation at 5 keV generated M-centre defects, producing an absorption feature around 720 nm, as well as a very small shoulder in the 580 nm region, indicating the potential presence of colloids of Na. Both spectra, 10 and 30 wt% concentrations, show similar F-centre, Na colloids and M-centre intensities. As described in section 2.1, inter-mixtures behave very differently from the salty ices particles (intra-mixtures) as the NaCl grains in inter-mixtures are not linked to the water matrix and are rather reacting to radiation in a similar way to pure NaCl grains. Indeed, we find a good agreement between the results coming from this experiment and the results of irradiations performed on pure NaCl grains by (Hand and Carlson, 2015; Poston et al., 2017).

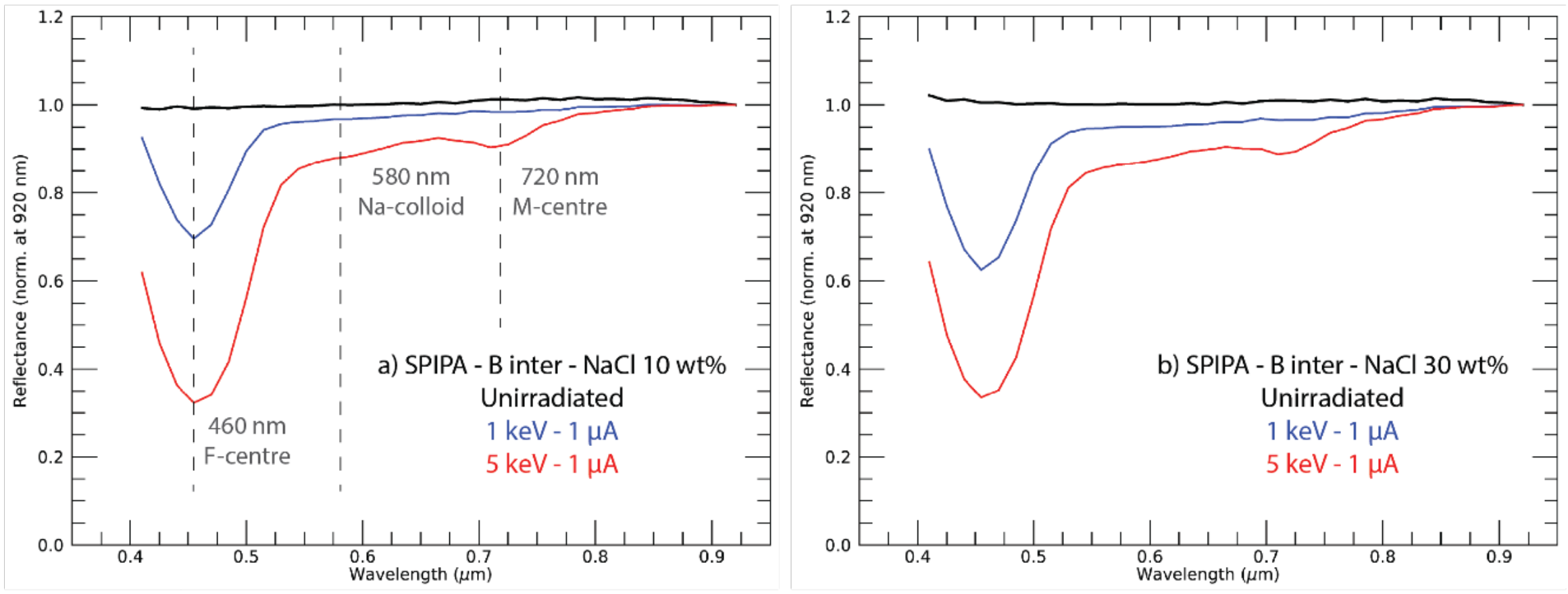

*Figure 4: Reflectance spectra of unirradiated and irradiated SPIPA-B / NaCl inter-mixtures with NaCl grains sieved to < 100 µm. Two concentrations of NaCl are compared: a) 10 wt% and b) 30 wt%. The wavelength positions of the crystal defects are shown in a). The black curves correspond to the non-irradiated samples. The blue curves correspond to the first irradiation performed on the sample at 1keV and 1 µA. The red curves are the results of the second irradiation performed at 5keV and 1µA. All spectra are normalised to their reflectance value at 920 nm to enhance the effect of irradiation visually.*

### 3.3.2. Intra mixtures of SPIPA-A ices

Figure 5 regroups the spectra of salty SPIPA-A ice samples before and after their irradiation. For all samples with a NaCl concentration of 5 wt%, the first irradiation produces a broad absorption complex over the entire visible spectral range. Within this broad complex, the two specific absorption features of the F-centre at 460 nm and the M-centre at 720 nm are recognisable as well as a maximum of absorption around 580 nm – at least for samples irradiated at 5 keV - but the overlap between these features makes them difficult to locate precisely.

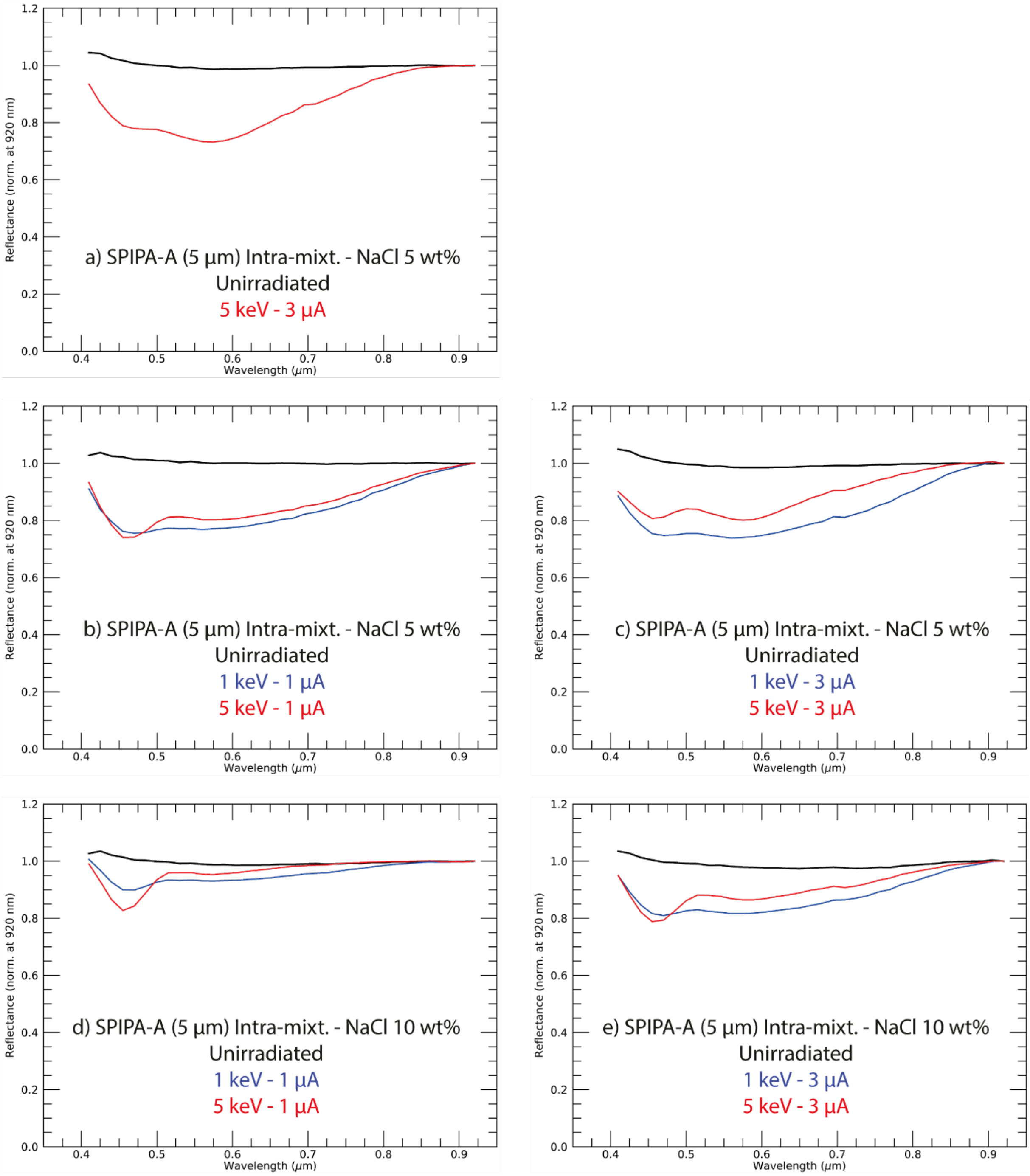

*Figure 5: Reflectance spectra of SPIPA-A granular ice samples with two different NaCl concentrations before and after irradiation. The black curves correspond to the samples before irradiation. The blue curves correspond to the first irradiation performed on the sample at energy and beam current specified in the legend on the plot. The red curves correspond to the second irradiation performed with electrons at higher energy. All spectra are normalised to their reflectance value at 920 nm to enhance the effect of irradiation visually. The SPIPA-A – 5wt% (a) is also performed on pristine ice but is displayed in red to note the higher energy and electron fluxes compared to the other cases.*

After the secondary irradiation, the absorption features already observed after the first irradiation were all more pronounced. The second irradiation generates a stronger F-centre with 5 keV electrons and with a lower electron flux (Figure 5, *b* and *d*) without significant modification to the rest of the spectrum; the 580-nm feature seems better defined for the SPIPA A sample with 10 wt% of NaCl after the secondary irradiation. With stronger electron fluxes (Figure 5, *c* and *e*), all bands are more pronounced. The 580-nm feature is comparable or even stronger than the F-centre with lower salt concentration (*a* and *c* compared to *e*, Figure 5). As a general trend in these experiments and perhaps counter-intuitively, the visible spectra of samples with the lower NaCl concentration (5wt%) appear more affected by the irradiation than spectra of sample with higher NaCl concentration (10wt%).

### 3.3.3. Intra mixtures of SPIPA-B ices

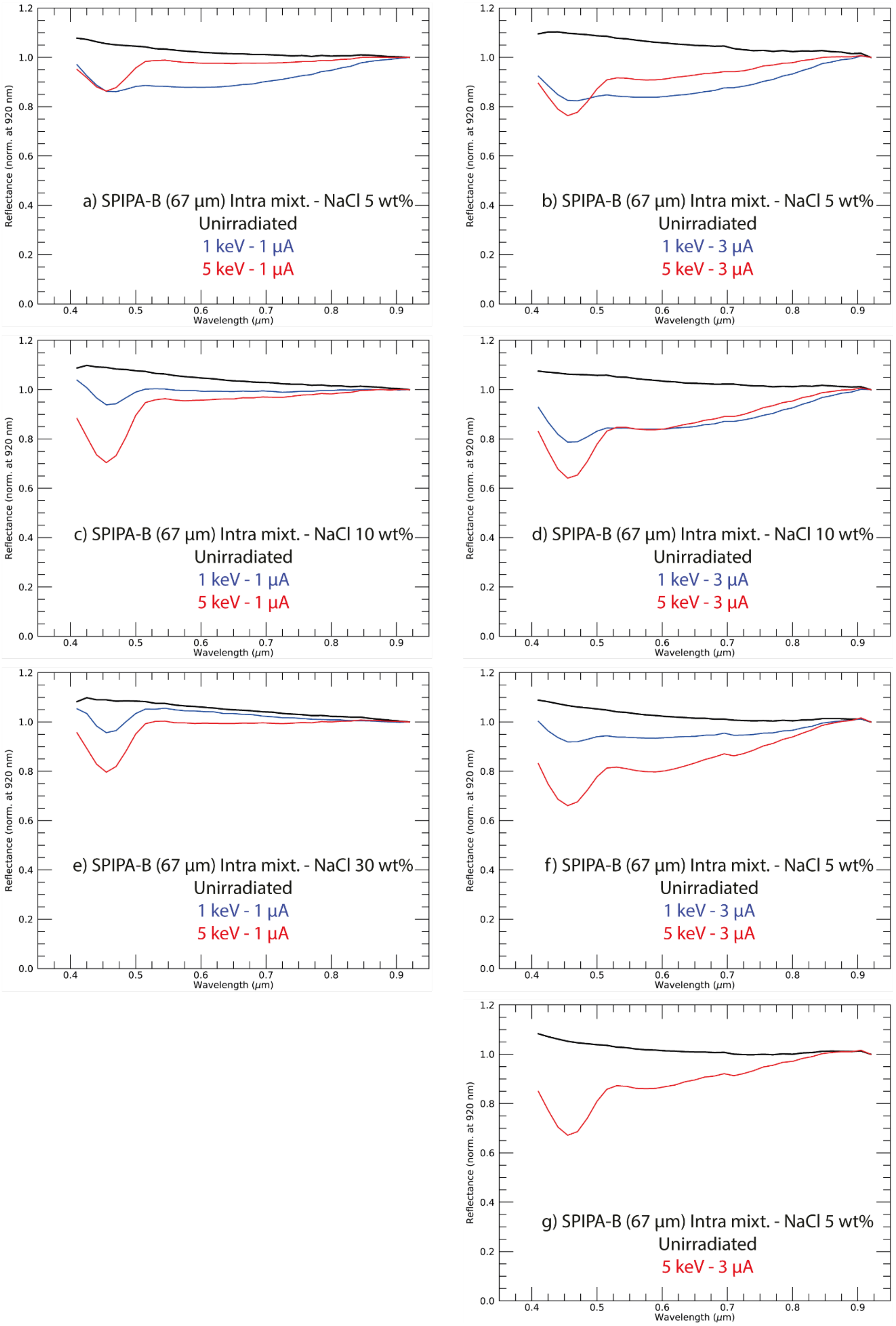

*Figure 6: Reflectance spectra of unirradiated and irradiated NaCl SPIPA-B granular ice samples with three different NaCl concentrations before and after irradiations. The black curves correspond to the samples before irradiations. The blue curves correspond to the first irradiation performed on the sample at energy and beam*

*current specified in the legend on the plot. The red curves correspond to the second irradiation performed with electrons at higher energy. All spectra are normalised to their reflectance value at 920 nm to enhance the effect of irradiation visually. The left panel of plots represents the irradiations performed with the low current (low electron fluxes). The right panel of plots represents the irradiations done with higher electrons fluxes. f) and g) are two different samples of SPIPA-B from Figure 2, B).*

Figure 6 regroups the VIS spectra of SPIPA-B granular ice samples with three concentrations of NaCl – 5, 10 and 30 wt% - before and after irradiation.

The samples irradiated with the lower electron flux (Figure 6, *a*, *c* and *e*) exhibit the formation of the F-centre for NaCl concentrations of 10 and 30 wt%. With 5 wt% NaCl, a broad complex of absorption over the entire VIS range is observed. It suggests the potential role of Na colloids, responsible for the maximal absorption at 580 nm, the 720-nm feature of the M-centre and possibly even another feature resulting from irradiation with electrons on NaCl crystals, such as the $F_4$ defect, centred at 825 nm (Schwartz et al., 2008). When the energy of the electrons increases, the F-centre becomes more pronounced for 10 and 30 wt% samples. For the 5 wt% sample, the F-centre remains of similar intensity, but the 580-nm and 720-nm features are no longer observed nor suggested.

Irradiation of samples with higher electrons fluxes (*b*, *d* and *f in* Figure 6) produced an absorption complex in the VIS (as a combination of the absorptions generated by the F- and M-centres and Na-colloids) even for low concentrations of NaCl in ice. The *b)* and *f)* plots in Figure 6 are replicas for the same sample and same irradiation conditions. They exhibit the same spectral pattern with a pronounced F-centre, but the strength of the absorption complex within the VIS varies. The secondary irradiations at higher energy have a similar effect on the samples, with a significant deepening of the F-centre. The absorption at 580 nm related to Na colloids is of similar intensity for both 5 and 10 wt% concentrations of NaCl. Irradiation at higher energy directly onto pristine salty ice (*g*, Figure 6) has generated similar absorption features as the secondary irradiation on the same sample (*f*, Figure 6) and similar to the same irradiation performed on another sample (*b*, Figure 6). They all exhibit a dominant F-centre, a weak but nonetheless marked absorption feature at 580 nm as well as a 720 nm feature.

## 3.4. Irradiation of slabs

Figure 7 presents the spectral results of the irradiations performed on compact slab ice samples. Irrelevant of the energy and flux of the electrons used to irradiate the slabs, the spectra are characterised by the presence of the one and only F-centre absorption feature.

Irradiation with a lower electron flux generates the F-centre feature and results in a negative correlation between the salt concentration and the intensity of the F-centre absorption. The spectra of the slabs with 20 and 30 wt% NaCl irradiated under a low electron flux (*f* and *d*, Figure 7) appear very similar, and both display significantly weaker features than the spectrum of the 10 wt% NaCl sample produced in the same conditions. With higher energies, the F-centre absorption deepens. The strongest absorption band is observed in the sample with 10 wt% NaCl.

Under irradiation with higher electron fluxes (*a*, *c* and *e* Figure 7), a linear trend is observed: the intensity of the F-centre correlates negatively with the amount of salt. When increasing the energy of the electrons, the F-centre deepens, and the negative correlation is maintained.

The first irradiation with a higher current (hence a higher flux of electrons) generates a stronger F-centre than the first irradiation conducted with a lower current. The intensity of the difference between these two first irradiation runs seems to be correlated with the salt concentration, being a maximum for the slab with 10 wt% NaCl.

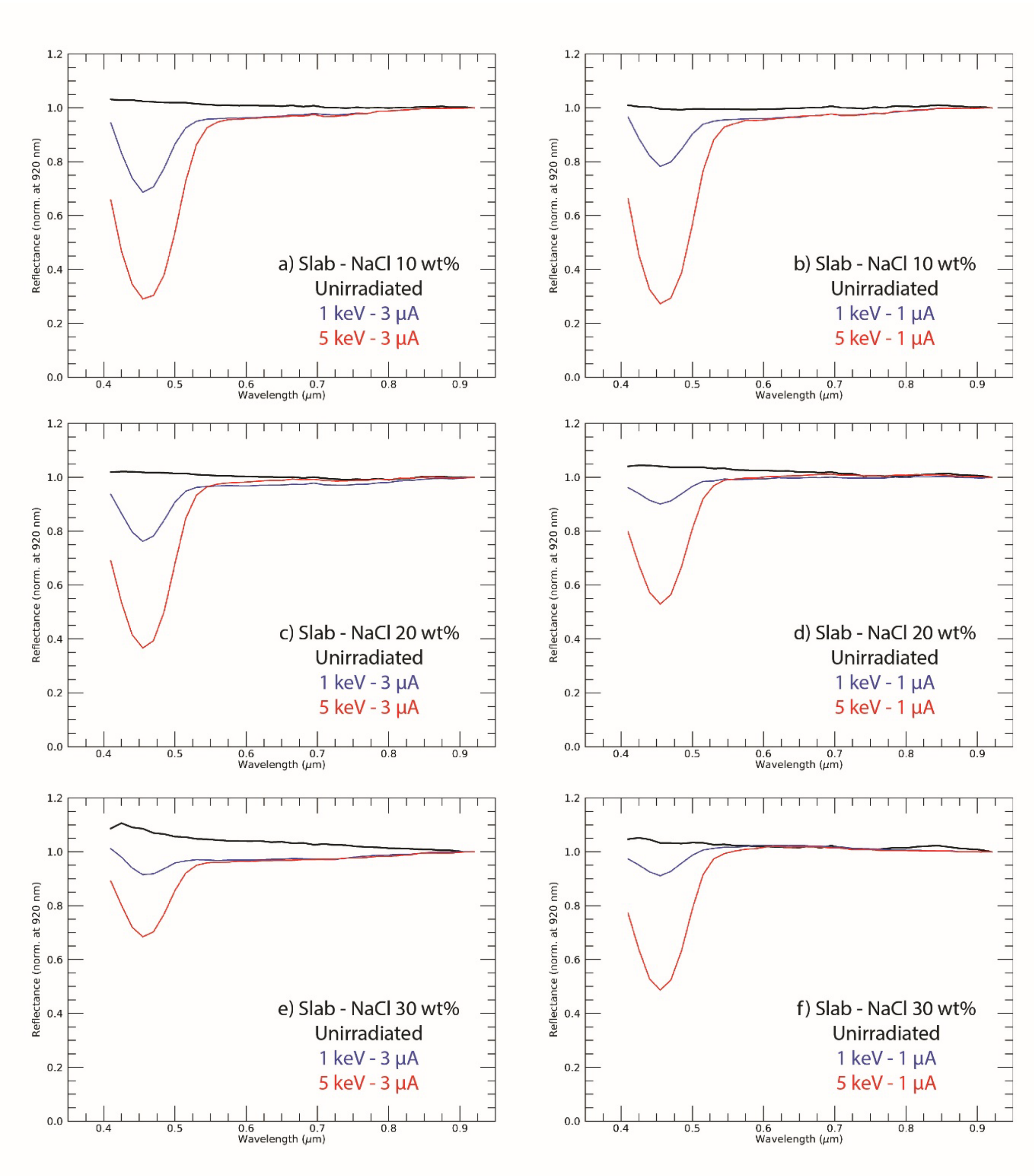

*Figure 7: Reflectance spectra of salty slabs with three different NaCl concentrations before and after irradiations. The black curves correspond to the samples before irradiation. The blue curves correspond to the first irradiation performed on the sample at energy and beam current specified in the legend on the plot. The red curves correspond to the second irradiation performed with electrons at higher energy. All spectra are normalised to their reflectance value at 920 nm to enhance the effect of irradiation visually. The left plots regroup the irradiation with higher electron fluxes, at 3 µA; the right plots regroup the sample irradiated with lower electron fluxes at 1 µA.*

## 3.5. Stronger irradiations on slabs

Figure 8 shows supplementary irradiations performed at 5 keV and 5 µA on slabs with 10 and 30 wt% NaCl after the irradiations at 5 keV and 3 µA. The purpose of this additional experiment was to simulate a longer period of irradiation and observe if the M-centre would finally appear on the slab samples. The blue and green spectra have both been irradiated with electrons fluxes in the order of $3{,}0 \cdot 10^{17}$ $e^{-}/cm^{2}$, which is three times more than the secondary irradiations. Consequently, a weak absorption feature is identifiable in the 10 wt% spectrum, but still absent in the 30 wt% case.

Taking into account all uncertainties, the third irradiation of the slab with 10 wt% NaCl does not create a stronger F-centre than the one formed during the secondary irradiation. This observation indicates that there is a range of crystal disorder generated by the secondary irradiation, which encompasses the effects of the ternary irradiation. The difference between the intensities of the F-centres is at a maximum between the 10 and 30 wt% NaCl samples.

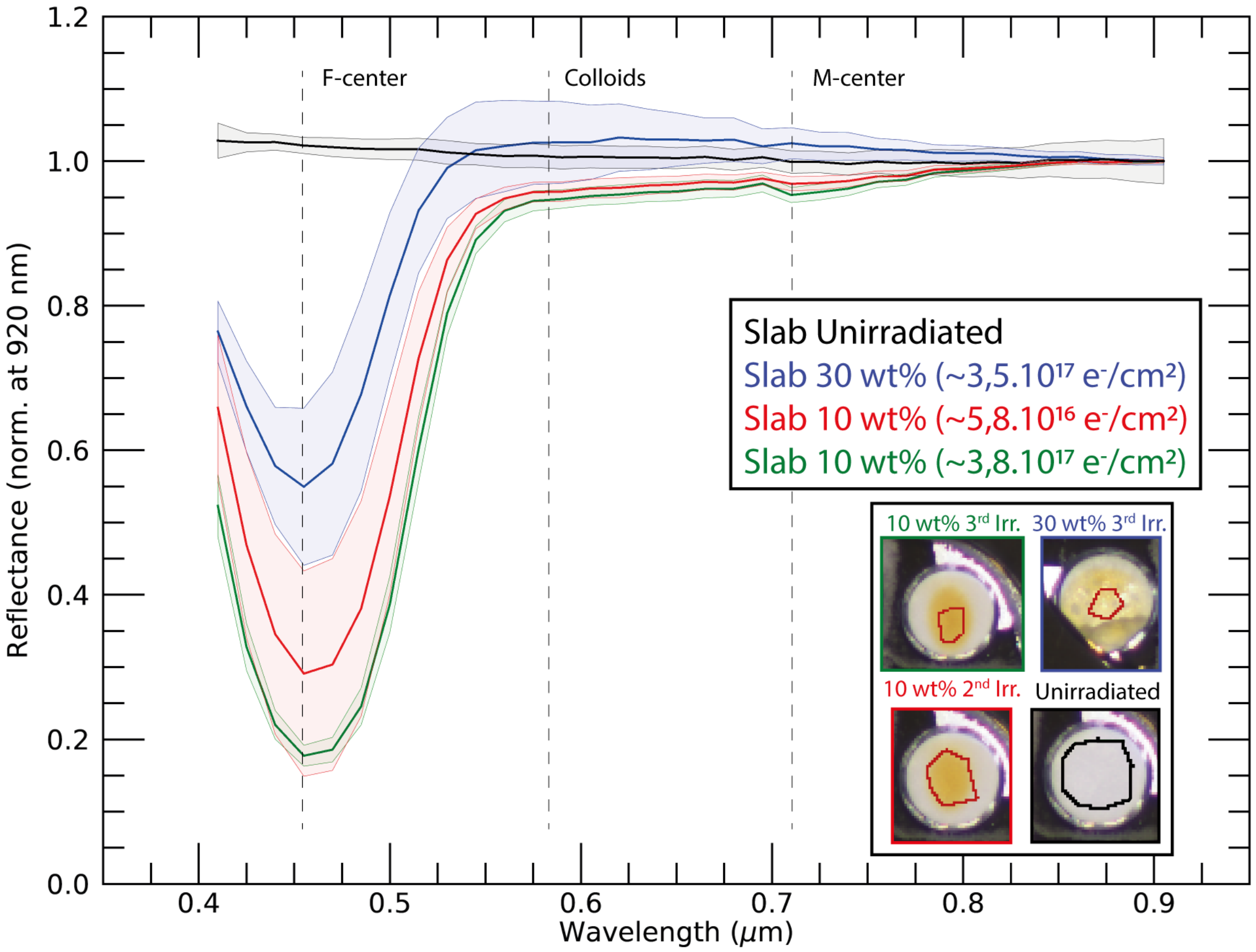

*Figure 8: VIS spectra of slabs irradiated with higher beam current (hence higher electron flux) than in Figure 7. The red spectrum corresponds to the secondary irradiation performed on the slab with 10 wt% NaCl from (red curve on b), Figure 7). The green spectrum corresponds to the same sample irradiated with a beam current of 5 µA. The blue spectrum is the sample from (slab e), Figure 7) after ternary irradiation performed at 5 µA. The four pictures in the lower-right corner are RGB colour composites of the samples, the 10 and 30 wt% are the same samples as the ones shown in D), Figure 2 but after the ternary irradiation. The red and black polygons on the pictures represent the ROIs defined for the averaging.*

# 4. Discussion

## 4.1. Quantifying the yellow colouration of salty ice samples

During the irradiation experiments, the formation of colour centres has been observed visually and characterised spectrally. The colour centres manifest themselves as absorption bands positioned at specific wavelengths, as described by (Doyle, 1958; Hand and Carlson, 2015; Hibbitts et al., 2019; Poston et al., 2017; Schwartz et al., 2008; Seitz, 1946). We also report on the absorption feature located at 580 nm, which we attribute to the formation of Na colloids during our experiments. We note that this absorption had not been documented at cryogenic temperatures in previous studies (Carlson et al., 2005; Poston et al., 2017). In order to identify trends and patterns associated with the behaviour of the different analogues, we compute the band depths of the absorption features at 460, 580 and 720 nm (at 455, 575 and 710 nm exactly with our spectral sampling). Similar computations have already been performed in the study of (Poston et al., 2017) in order to derive the age of young geological features affected by irradiation. Subsections 4.1.1 to 4.1.3 focus on the formation process of the colouration centres and crystal defects. The comparison of granular ices and compact slabs is presented in section 4.2. The roles of grain size and salt concentration are discussed in section 4.3 and 4.4, respectively.

### 4.1.1. Band depth at 460 nm, F - centre formation

The values of bands depth for SPIPA-A and SPIPA-B particles are in the range of 10 to 40% depending on the energy of the electrons and the total electron deposition. For SPIPA-B samples, we noted an increase in the value of band depth for a given current when changing the energy of the electrons (Figure 11). This tendency is not as pronounced for SPIPA-A samples. For SPIPA-A samples, the intensity of the F-centre does not evolve significantly between the first and the second irradiation, but it does for SPIPA-B samples (Figure 5, Figure 6).

Compact slabs appear more affected than granular ices by variations in the energy of the electrons. There is a clear increase in the intensity of the band depths when comparing the results of irradiation at a given current but at either 1 or 5 keV. There is a factor of 2 or more in the value of the band depth of irradiated slabs (comparing 1 and 5keV), which is maintained even at higher electron fluxes. Interestingly, Irradiations performed with constant energy of 5 keV but at a current of 1 or 3 µA generate similar band depths.

The difference observed between slabs’ band depths (green squares, Figure 11) and granular particles’ band depths irradiated with higher electron fluxes (yellow circles and triangles, Figure 11) is significant. It confirms that the formation of stronger F-centres with slabs is a feature.

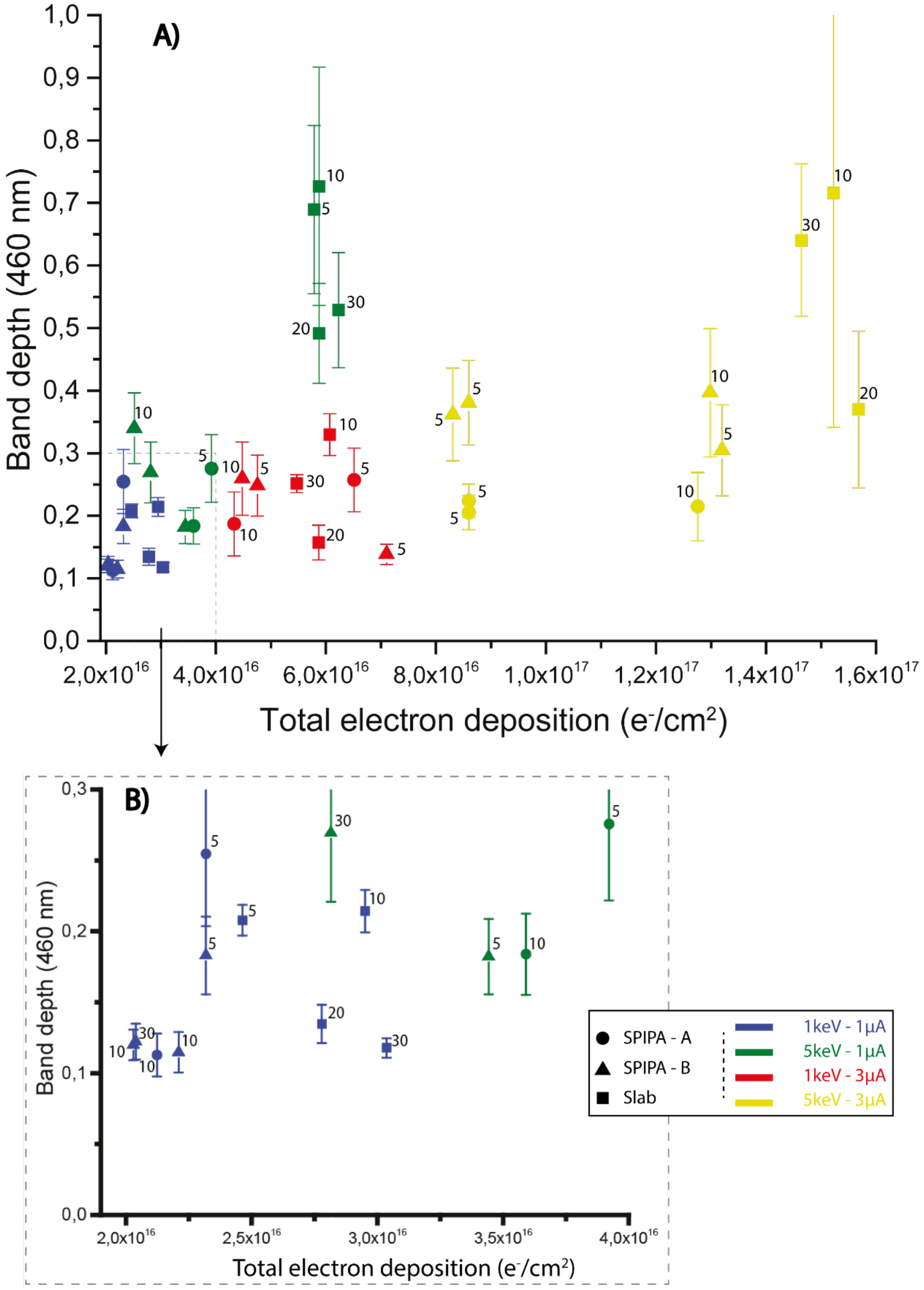

*Figure 9: 460-nm band depth (computed at 455 nm). The triangles and the circles represent SPIPA-B and SPIPA-A samples, respectively. The blue points represent the irradiations at 1 keV - 1 μA; Green: 5 keV – 1 μA; Red: 1keV - 3 μA; Yellow: 5 keV - 3 μA. The number close to each point indicates the NaCl concentration (wt. %). B) is an enlargement of the dotted region on A).*

## 4.1.2. Band depth at 580 nm, aggregates of sodium colloids

The production of sodium colloids has been studied previously, primarily in the field of storage of highly radioactive waste (Soppe et al., 1994; Sugonyako, 2007; Swyler et al., 1980). Their formation process is intrinsically linked to the formation of the F-centre, which results from the filling of an anion vacancy by an electron (Pooley, 1966). The first stage is the implantation of an exciton (electron – hole pair) in the crystal lattice. It leads to, by relaxation of the exciton's energy within the lattice, the "migration" of the $Cl^-$ from one site to a neighbouring site, thus forming a $Cl_2^-$ molecular ion (which is equivalent to an interstitial $Cl^0$ atom with a covalent bond to another neighbouring $Cl^-$ site). During this process, an electron is trapped in a vacant $Cl^-$ site, forming the F-centre and the interstitial $Cl^0$ atom is referred to as the H-centre (Hartog et al., 1994; Soppe et al., 1994). This H-centre does not produce any absorption in the visible spectral range.

The F-centre is a mobile defect within the lattice. It moves due to "jumping" of a $Cl^-$ neighbour into the vacancy. However, this jump requires a high amount of energy and is, therefore, strongly linked to the dose rate as well as the temperature. If several F-centres gather through this mechanism, it creates a local cluster of F-centre defects, leading to the formation of localised colloids of sodium (Hartog et al., 1994; Sugonyako, 2007).

Figure 10 shows the band depths at 580 nm, computed here at 575 nm due to our wavelength sampling, versus the electron flux for the two types of granular ices. The slabs are not presented in this figure because, as seen in section 3.4, they do not produce any other colour centre than the F-centre, as further discussed in section 4.2.

The spectral results of the production of sodium colloids appears higher with a lower concentration of salt. Following irradiation at 1 keV and 1 µA, the band depths are stronger for the 5 wt% samples than for the 10 wt% NaCl. This trend is not unambiguously maintained for the other irradiations, as some 10 wt% samples occasionally show values of band depths comparable to 5wt% samples. The highest values of band depths are nevertheless observed within 5 wt% samples.

Irradiation experiments previously performed on pure NaCl grains have not always resulted in observations of features at 580 nm. For instance, (Poston et al., 2017) have not identified any 580-nm feature on pure NaCl grains during their irradiations. (Hand and Carlson, 2015) have noticed the possible presence of this feature in pure NaCl grains as well as a weaker feature in their analogue consisting of NaCl grains covered by $H_2O$ ice. The explanation proposed by (Poston et al., 2017) and based on previous studies (Schwartz et al., 2008; Sugonyako, 2007; Swyler et al., 1980) involves the rate of energetic particles deposition (dose rate). In their work, the electron fluxes were lower (250 nA maximum) than in the present study (1 and 3 µA) where we observed almost systematically the formation of a 580-nm feature in granular ices.

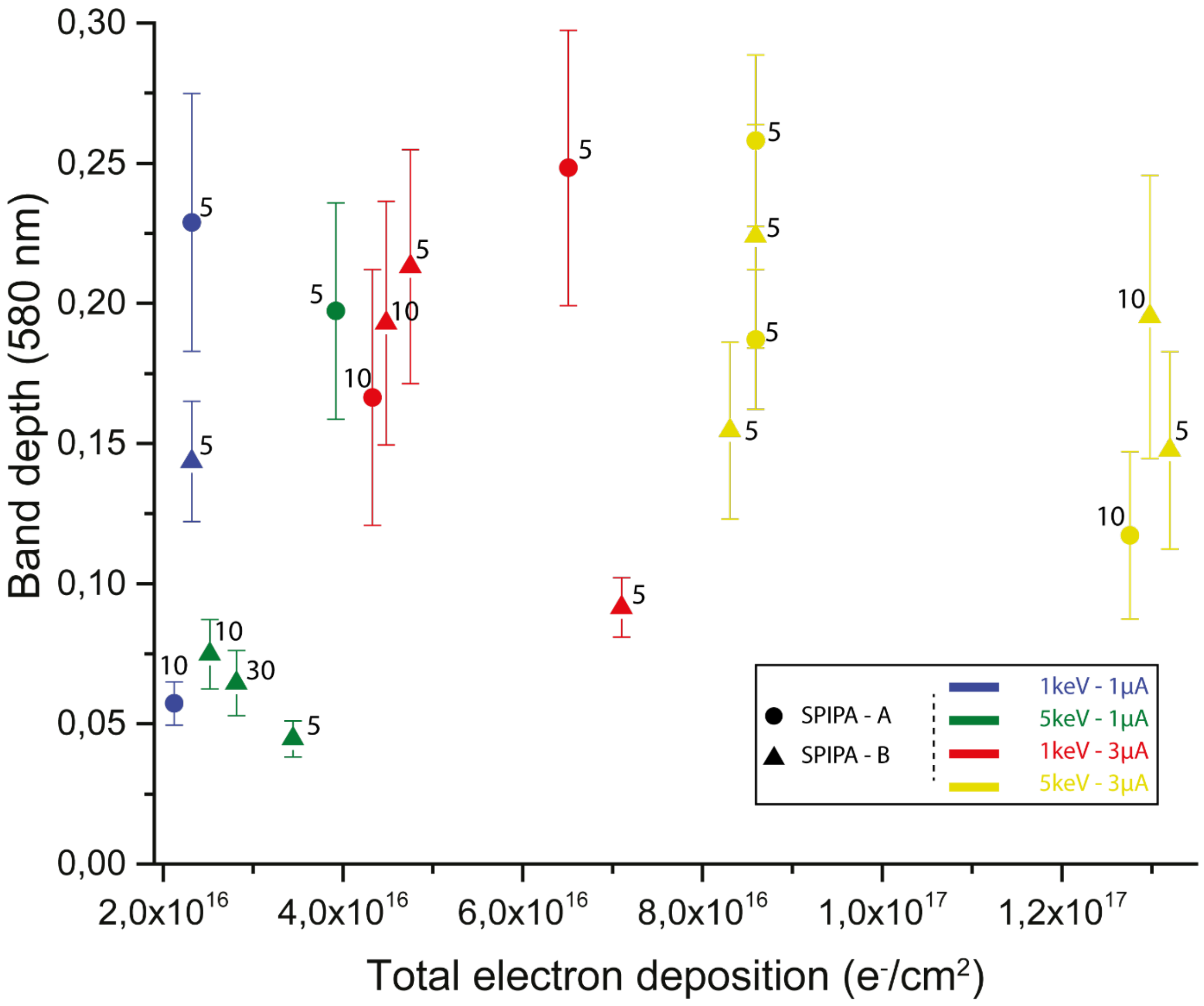

*Figure 10: 580-nm band depth (computed here at 575 nm). The triangles and the circles represent the SPIPA-B and SPIPA-A samples, respectively. The blue points represent the irradiations at 1 keV - 1 µA; Green: 5 keV – 1 µA; Red: 1keV - 3 µA; Yellow: 5 keV - 3 µA. Slabs are not represented here because absorptions at 580 nm have not been identified for these samples. The number close to each point indicates the NaCl concentration (wt. %).*

### 4.1.3. Band depth at 720 nm, M - centre formation

Figure 10 regroups the band depths computed for the M-centre, which is formed when two electrons are trapped within a $Cl^-$ vacancy. This defect has never been observed during our experiments with compact slabs and is only seen in granular ice samples with relatively low concentrations of NaCl. It has to be mentioned, however, that our dataset contains more samples with low NaCl concentration as the production of fine-grained granular ice (SPIPA-A) is only possible with low salt concentration. Complementary measurements with higher concentrations of salts would provide better statistics and are desirable in the future.

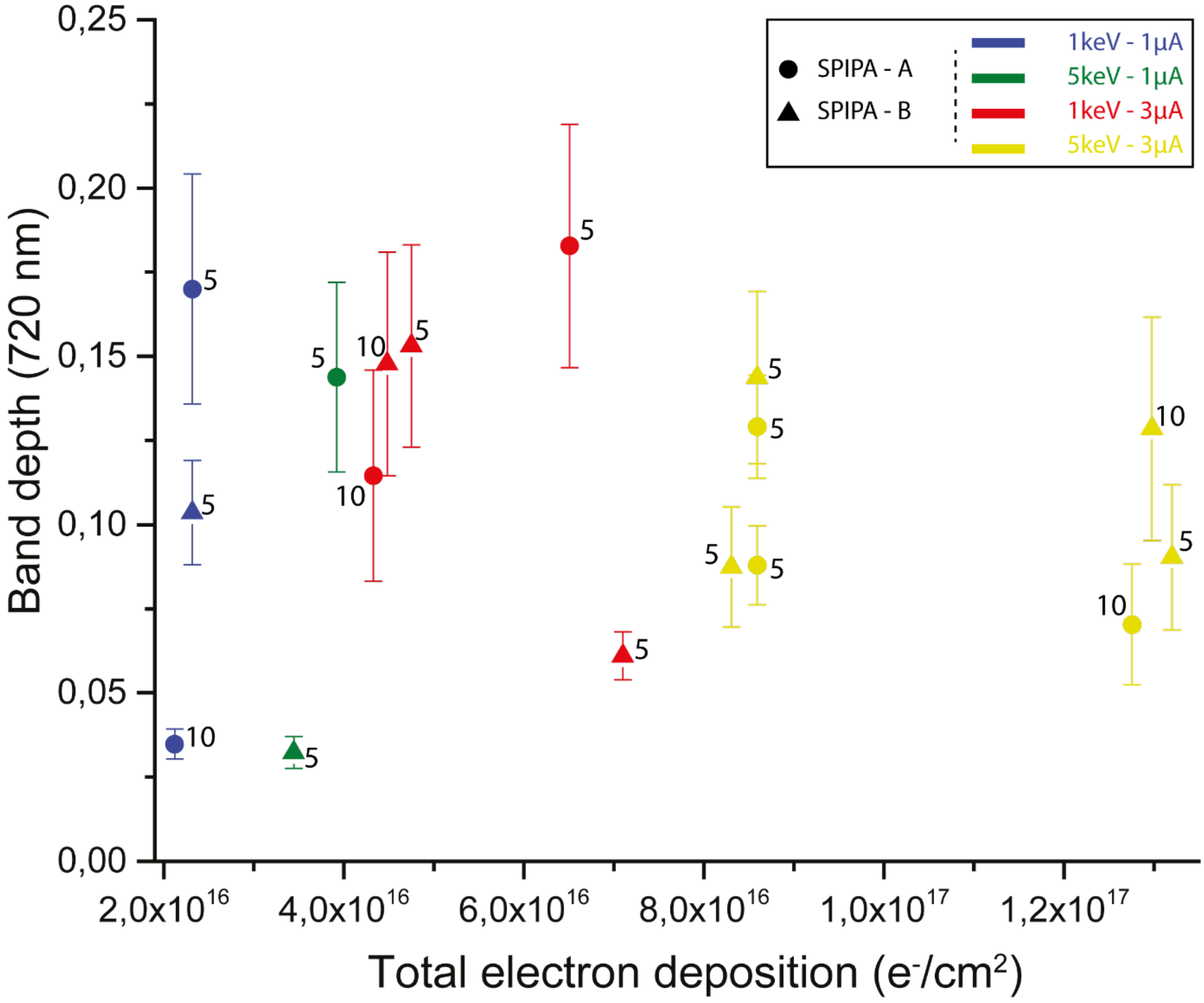

*Figure 11: 720-nm band depth for the different analogues (computed at 710 nm here). The triangles and the circles represent the SPIPA-B and SPIPA-A samples, respectively. The blue points represent the irradiations at 1 keV - 1 µA; Green: 5 keV – 1 µA; Red: 1keV - 3 µA; Yellow: 5 keV - 3 µA. Slabs are not represented here because absorptions at 580 nm have not been identified for these samples. The number close to each point indicates the NaCl concentration (wt. %).*

As with other spectral features, when the M-centre absorption is present, its maximum band depth is observed within the less concentrated samples. In most cases, the M-centre is not expressed as a strong absorption band but rather as a shoulder within a broad absorption complex over the entire VIS range or as a shallow absorption feature. These features suggest nevertheless that this colour centre is present, contributing significantly to the broad VIS-NIR complex, by extending it toward longer wavelengths up to 920 nm. The strongest feature has been observed with the 5 wt% SPIPA-A sample irradiated with 1 keV – 3 µA (*b* and *c*, Figure 5), although taking error bars into account, the 5 and 10 wt% SPIPA-B samples irradiated with 1 keV – 1 and 3 µA (b and d, Figure 6) or the 5 wt% SPIPA-A irradiated with 5 keV – 1 µA (b in Figure 5) can be comparable.

## 4.2. Differences between granular ices and slabs

The two types of analogues produced in this study induced a particular arrangement of the different components inside the ice particles (water and NaCl, as crystals, hydrates or glassy state). From our near-infrared spectroscopic study of the two types of samples (Cerubini et al., 2021), we have shown that particulate "flash-frozen" samples contain salts essentially in the form of anhydrous crystals (here NaCl, halite) whereas the compact slabs are dominated by the hydrated forms, in this case, the dihydrate $NaCl{\cdot}2H_2O$ (hydrohalite). Because of the high cooling rate used for their production, flash-frozen particles contain up to four different components: crystalline water ice ($I_c$, $I_h$), NaCl crystals, $NaCl{\cdot}2H_2O$ crystals and a glassy form of NaCl aqueous solution (Bove et al., 2015; Cerubini et al., 2021; Ludl et al., 2017). Compact slabs, on the other hand, are mainly made of hydrohalite and water ice both well crystallised.

Electron doses in our experiments are similar to those used in previous works (Hand and Carlson, 2015; Poston et al., 2017) but the 580-nm band is clearly recognisable in the case of our salty ice particles, while it was only suggested in previous studies (Fig.2 in (Hand and Carlson, 2015)). This feature is not present in all the samples tested so far. For instance, our compact slabs of salt and ice do not display any Na colloid absorption band with the irradiations pursued. Even after irradiation with higher doses (Figure 7 and Figure 8), slabs did not exhibit Na colloid formation.

The appearance of the M-centres in compact slabs was observable only after irradiation with higher doses (5 keV – 5 µA) specifically at 10 wt% concentration (Figure 8), while it has been observed more often in granular flash-frozen particles (Figure 5 and Figure 6). This observation is in agreement with the hypothesis that the formation of M-centres depends on the total dose of electrons received, rather than a particular process within the crystal lattice (Hartog et al., 1994; Schwartz et al., 2008; Sugonyako, 2007; Trumbo et al., 2019).

Electrons are responsible for most of the disorder within the crystal lattice. In our analogues, however, electrons are not systematically interacting with salt particles because of the presence of water molecules, which leads to the less efficient production of defects compare to pure salt crystals.

With similar energies and electron fluxes, the main conclusive difference observed with spectroscopy is the stronger F-centres for the compact slab samples, and the presence of colloid signatures only for granular samples. Our results show how important the production of icy analogues samples is in the scope of understanding the surface of icy moons. The presence of salt inside the ice but also the way it is mixed with water ice play key roles in their colouration behaviour.

### 4.2.1. Na colloids production in granular samples

The presence of a broad absorption at 580 nm is caused by the resonance bands related to aggregates of Na colloids, as previously mentioned by (Doyle, 1958; Hand and Carlson, 2015; Poston et al., 2017; Schwartz et al., 2008; Sugonyako, 2007). In previous studies, pure NaCl crystals have been irradiated. Only (Hand and Carlson, 2015) have investigated the behaviour of samples of NaCl grains covered with water ice particles, or used a frozen brine saturated with NaCl.

Na colloids and the resulting absorption at 580 nm have been observed in our analogues, albeit only in the case of particulate samples (Figure 5, Figure 6 and Figure 7).

The higher amount of hydrates in slabs (see section 4.2), compared to granular flash-frozen particles, may necessitate higher energies to initiate the “swapping” between an F-centre and a $Cl^-$ anion, therefore reducing the efficiency of irradiation to create Na colloids. If the dehydration is indeed mandatory before colour centres can be generated as suggested by (Thomas et al., 2017), then the Na colloids could appear under stronger and/or longer irradiation, as hydrates would lose water to form anhydrous NaCl. Since we were able to form Na colloids in granular salty ices with our current dose rate, we should also be able to form the same defect in pure NaCl grains. The electrons would affect only NaCl crystals instead of, in the case of our analogues, salt crystals possibly hydrated, embedded within a water ice matrix.

Intrinsically linked to the formation of F-centre, the formation and mobility of H-centres also plays an indirect role in the growth rate of Na colloids. As previously explained (Section 4.1.1), the F-centre is a mobile defect, but the H-centre is even more mobile, as it moves following the bonds within the crystal lattice and therefore requires a lower amount of energy to be displaced (Hartog et al., 1994; Sugonyako, 2007). The two types of defects can move, and F-centres can aggregate to locally form Na colloids. The F- and H-centres can also recombine and annihilate each other. The formation of clusters of defects will lead to the growth of Na colloids absorption but is dependent on the production rate of F- and H-centres, which has to be higher than their recombination rate (Sugonyako, 2007). Therefore, it depends on the mobility of the defects, which in turn depends on the temperature and on the dose rate. The compact slabs are more efficiently cooled-down under vacuum conditions compared to granular particles, therefore reducing the mobility of crystal defects (Schwartz et al., 2008; Sugonyako, 2007; Swyler et al., 1980). It eventually affects the displacement of F-centres and ultimately the formation of Na colloid.

### 4.2.2. Intensity of absorption by the F-centres

The absorption band of F-centres is more pronounced in compact slabs than in granular samples. The more effective cooling of compact slabs under vacuum conditions limits the crystal defect mobility as mentioned in section 4.1.1 and 4.2.1. Aggregation of F-centres as well as recombination, if any, with H-centres, is then reduced compared to granular samples and favour the conservation of F-centres. Increasing the energy of the electrons results in interactions at larger depth within the sample (Galli et al., 2017; Hand and Carlson, 2011). Therefore, the depth scanned with spectroscopy samples higher proportion of crystal defects. It results in a nearly linear positive correlation between the energy and the absorption feature within slab samples (given a certain current).

In granular samples, the formation of Na colloids reduces the amount of F-centres, as being the result of the displacement and aggregation of F-centres. This is the result of higher mobility of the F-centres within granular samples (section 4.1.1).

With stronger and longer irradiation of granular samples, the evolution of band depths (at all wavelengths) never exhibited a strong positive correlation with total electron deposition. We suggest that salty granular ice particles reach a dynamic equilibrium between production and recombination. The amount of Na colloids is impacted by the migration of H-centres, as the recombination of F-H pairs limits, in the end, the amount of available F-centres to aggregate.

To conclude this comparison between the spectral behaviours of compact slabs and granular salty ices, it has to be mentioned that the effects of the dose rate and total dose deposition on crystal lattice under electron irradiation are poorly constrained at cryogenic temperatures, even with pure NaCl crystals. Our study involves NaCl grains associated in different ways to a water ice matrix, therefore adding additional complexity to this system. More dedicated analyses of the fine structure of our analogues, such as those used in (Hartog et al., 1994; Schwartz et al., 2008; Sugonyako, 2007) would be very useful to better constrain the properties of the crystals. As our analogues are icy samples, it would require further precautions and to some extent prevent some analysis from being realised (as temperature rise over the eutectic point). Nevertheless, we saw that specific absorption features are related to a specific ice production procedure, which can be useful to better constrain the composition of the surface of Europa with VIS spectroscopy. This could be applied to future space mission results to help understanding the structure and composition of icy surfaces.

## 4.3. Role of the grain size

In particulate surfaces, grain size is among the most important parameters which controls the absolute amount of reflected light and the intensity of the spectral features. The situation is complicated here as the sizes of both the ice and the salt particles play independent and major roles.

Indeed, for most of our samples, light is scattered by the ice particles and the light transmitted through the ice is again scattered by the smaller salt particles.

Systematic comparisons of the spectral behaviour of SPIPA-A and SPIPA-B analogues with the same NaCl concentrations that underwent the same electron irradiation highlight the influence of the ice grain size on the visible colouration induced by crystal lattice modifications in the salt. With larger ice grains, the total optical path length of the light within the ice is longer than with smaller ice grains, resulting in enhanced spectral features of the salt inclusions within the ice. Considering the different analogues produced in the present study, the effective ice scatterers sizes can be classified as follows: effective grain size for slabs ($GS_{slab}$) > $GS_{SpipaB}$ > $GS_{SpipaA}$ (Pommerol et al., 2019). Therefore, in the absence of other effects, we would observe a marked decrease in the strength of the absorption features following this order. The only common feature of every analogue is the F-centre. For similar irradiations, the slabs show stronger F-centres than the granular ices. Both the grain sizes of the ice and the salt crystals contribute to this effect. Indeed, we expect that slow crystallisation in the slabs results in large salt crystals in addition to long optical path length within the ice. Although SPIPA-A and –B methods produce different ice grain sizes, the size of the salt crystals inside these ice particles is possibly very similar, as both are produced by flash-freezing procedure, and certainly much smaller than in slabs.

The influence of ice particle size seems less pronounced overall when comparing the SPIPA-A and -B granular samples, in particular at low dose and energies. At higher dose and energy (5 keV – 3 µA, Figure 9), SPIPA-A samples display shallower bands than SPIPA-B samples, themselves having smaller band depths than slab samples. Other plausible reasons for the observed variability in the intensity of the F-centres between the different types of analogues have been proposed in section 4.2.2. The grain size also logically contributes to the observed differences between the analogues but is most likely not the sole explanation. Moreover, SPIPA-B inter-mixtures (Figure 4) show intensities of the M-centre relatively close to the ones seen in slabs, although the grain size of the salt was under 100 µm.

Ultimately, the porosity of the sample can also affect the efficiency of the electron irradiation to produce crystal defects. The SPIPA-A and –B ices have 80 and 49 % of porosity respectively (Pommerol et al., 2019), compared to the close to 0 % for slab samples.

## 4.4. Role of the amount of salt in the sample

The relation between band depth and salt concentration is rather counter-intuitive: for all samples, slabs and granular, the strongest absorption features are always seen with samples less concentrated in NaCl. The intensity of the band depth of the F-centre increases with the salt concentration up to 10 wt%, but then decreases with higher salt concentration (30 wt%), as seen in Figure 6 and Figure 7 for granular particles and compact slabs, respectively. An explanation to this behaviour will necessarily be complex as light scattering and production of electrons-induced defects occur at different depths and both processes are affected differentially by the salt concentration.

The penetration depth of the electrons within icy particles is rather small compared to the optical path length of photons. Electrons penetrate much less than 1 µm within the ice grains, depending on their exact energy, and therefore only affect their periphery, even for our smallest grains with diameters of 4.5 µm. Visible photons however are scattered and transmitted by the ice grains without absorption. According to (Hapke, 2012), the thickness of the scattering layer for such transparent samples is about 100 times the diameter of the particles divided by the porosity. This represents about 20 mm for pure SPIPA-B (67 µm) and a few millimetres for the very porous samples made of the small SPIPA-A (4.5 µm) particles. Our pure ice slabs are translucent and one can see the sample holder through the ice. The situation changes drastically however as small salt crystals are present into the ice in increasing number. The small salts crystals act as very efficient internal scatterers and will strongly reduce the penetration of photons within the sample, eventually preventing photons to be transmitted through the ice grains (Figure 12, b).

Another important consideration here is the limiting factor in the production of the crystal defects. At very low salt concentration, there must be a first regime where the quantity of salt crystals encountered by the electrons limits the production of the colour centres. In this case, the intensity of the colour centre must increase with salt concentration (from 5 to 10 wt.% in our work). As the salt concentration increases however, we will pass a point where it is not anymore the salt concentration that limits the production of colour centres, but the total dose and hence the number of deposited electrons. Evidences for such a regime comes from observations, in this work as well as in previous studies (Poston et al., 2017), that repeated irradiations of a sample at the same location result in a deepening of the absorption.

The complex relation observed between concentration and intensity of the spectral features is the result of the competition between the different processes mentioned. Figure 12 illustrates two cases to explain how a higher salt concentration can result in shallower absorptions. In Figure 12a, the salt concentration is low enough that photons are still transmitted through several ice grains and interact with a number of salt crystals either presenting defects (in red) or still pristine (in blue). In both cases, the total electron dose, not the salt concentration, limits the production of the colour centres and their absolute amount is similar. Their relative amount is however much higher in case 12a as the total amount of salt crystals is lower. The result is a dilution of the defects as the salt concentration increases. In Figure 12b, the salt concentration is such that all photons are backscattered within the volume of a single ice particle. At lower salt concentration (Figure 12a), photons are able to cross entire particles and will see proportionally more surfaces of ice particles where defects are located. As a result, interactions between photons and crystal defects are more numerous in 12a than 12b and the absorption features will be stronger.

While the hypothesis formulated here seems plausible and is supported by our observations, we acknowledge that this is only an idealised conceptual explanation and that a quantitative physical model or simulation will be necessary to verify several aspects and gain confidence in the explanation.

It will be a challenging task however, as both the light scattering and the photons interactions with the granular porous surface remain complex processes to simulate and they will need to be combined here. Finally, other effects could also play a role. For instance, adding salt within the ice increases the bulk density of the particles and therefore decreases the penetration depth of the electrons. This effect would be very small in bulk, if the salt particles are homogeneously distributed within the particles but could be much larger if the salt particles are locally concentrated, for instance in micro-cracks of the ice.

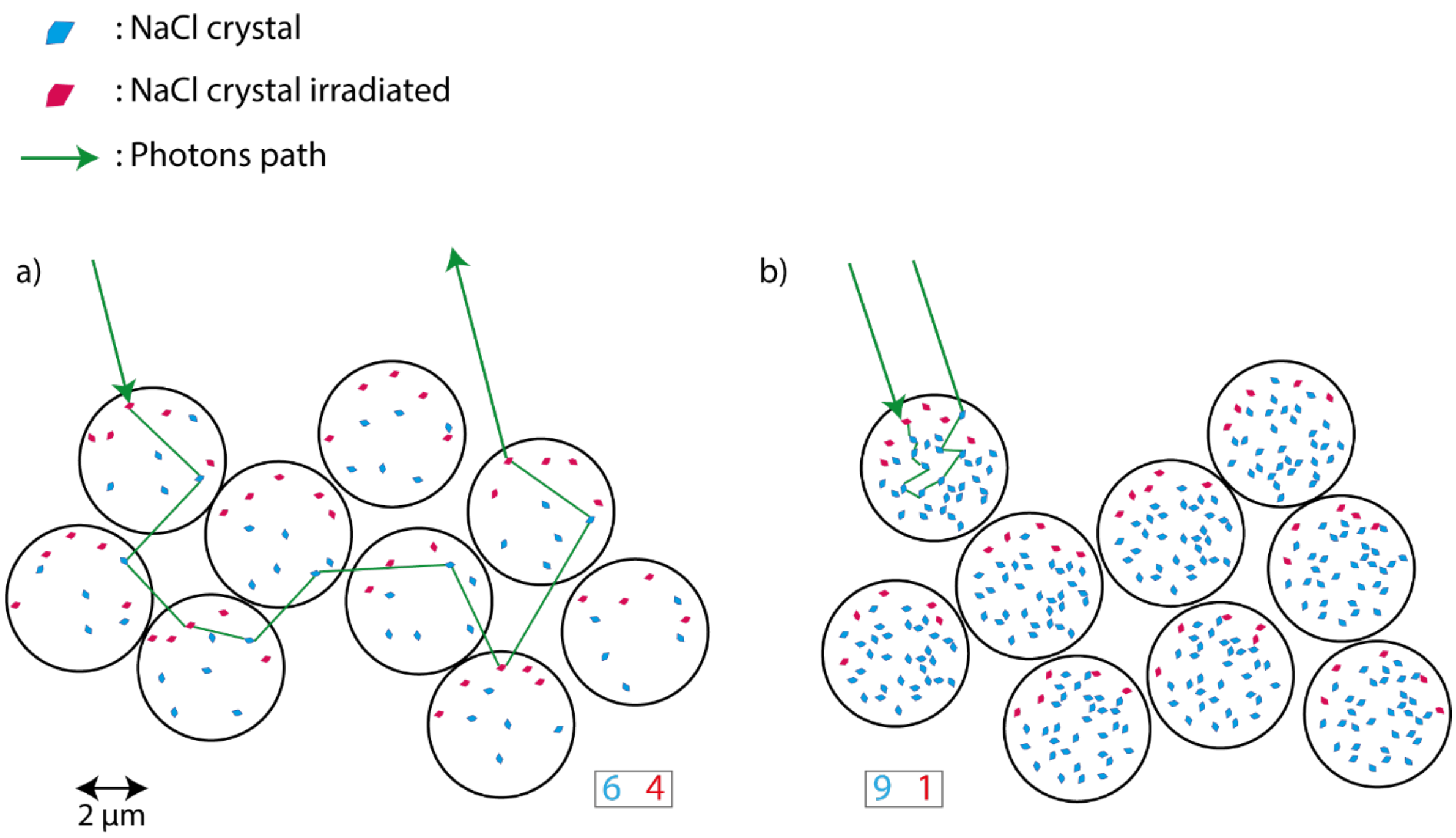

*Figure 12: Schematic representation of the path of visible light (photons) through salty ice particles with low (a) or high (b) salt concentration. The unirradiated salt crystal are indicated as blue particles. The irradiated and absorbing particles are indicated in red and are mostly found at the periphery of the grains as the penetration depth in ice (~1µm) of the electrons that produce these defects is smaller than the grain size. The green arrows represent the optical pathway of photons through the sample. At low salt concentration, photons can be transmitted through entire ice particles and can be multiply scattered between ice particles, increasing the likelihood of photons to encounter defects. At high salt concentration, most of the scattering takes place within the first particle of ice encountered, in its interior occupied by numerous pristine salt particles.*

## 4.5. Link to previous works and implications for icy moons

### 4.5.1. Comparison with previous studies

Figure 13 regroups all values of the ratio of band depths between F- and M-centres as defined by (Poston et al., 2017). We have seen the F-centre form with all types of analogues, with intensities comparable to previous studies. Our results differ from previous studies regarding the formation of M-centre, however. Unlike previous studies, we do not observe a regular growth of the M-centre with increasing irradiation. We rather observe the first phase with a growth of the F-centre followed by a second phase where the M-centre grows until saturation. With the electron energies and fluxes tested, we observed the M-centre mostly with granular ices, being favoured by either higher energies (5 keV), higher beam currents (3 µA), or the combination of both (Figure 5 and 6). The amount of salt is hypothesised to play a key role as all granular samples with 5 wt% concentration showed the formation of both F- and M- centres, even at low energies and currents. The M-centre is particularly difficult to produce with compact slab samples as it appears very weak and visible only under higher beam current (5 keV – 5 µA) for dose depositions of the order of $4{\cdot}10^{17}$ $e^-/cm^2$.

Our granular analogues show the production of a strong absorption feature related to the formation of colloids of Na as already suggested, but not observed, by (Hand and Carlson, 2015; Poston et al., 2017) at cryogenic temperatures with pure NaCl grains.

Our band depth ratios do not reveal obvious patterns. Considering each type of analogue separately, we see a common trend of decreasing ratios with increasing electron fluxes. This effect seems even stronger for SPIPA-B samples than for SPIPA-A samples. This behaviour is opposite to what (Poston et al., 2017) observed. This discrepancy is the result of the preferential formation of Na colloids instead of M-centres with our icy analogues. In most of our irradiation of granular samples, secondary irradiations enhance the F-centre intensity (460 nm), and the definition of the Na colloid feature (580 nm) but reduce the M-centre intensity (720 nm).

The *dose rate* we have used is slightly higher than the one used by (Hand and Carlson, 2015; Poston et al., 2017), which could be the reason for the formation of Na colloids in our samples, as discussed in section 4.2.1. On the other hand, the M-centres in our spectra appear less marked than in previous works. This could be related to the fact that the efficiency of irradiation to produce M-centres is higher with pure NaCl than with our analogues composed of a water ice matrix, hydrohalite, and pure NaCl crystals that are all potential targets that accommodate the excess energy provided by the electrons. The production of the M-centre being rather dose-dependent; its chances of production are reduced by the presence of other components.

The presence of salt within ice may also have an implication for the efficiency of the sputtering of water molecules or radiolytic products, therefore necessitating higher current and energies to disrupt the surface. Ion irradiation of salty ices has, to our knowledge, never been tested so far and may be of interest to better constrain exosphere – surface exchanges.

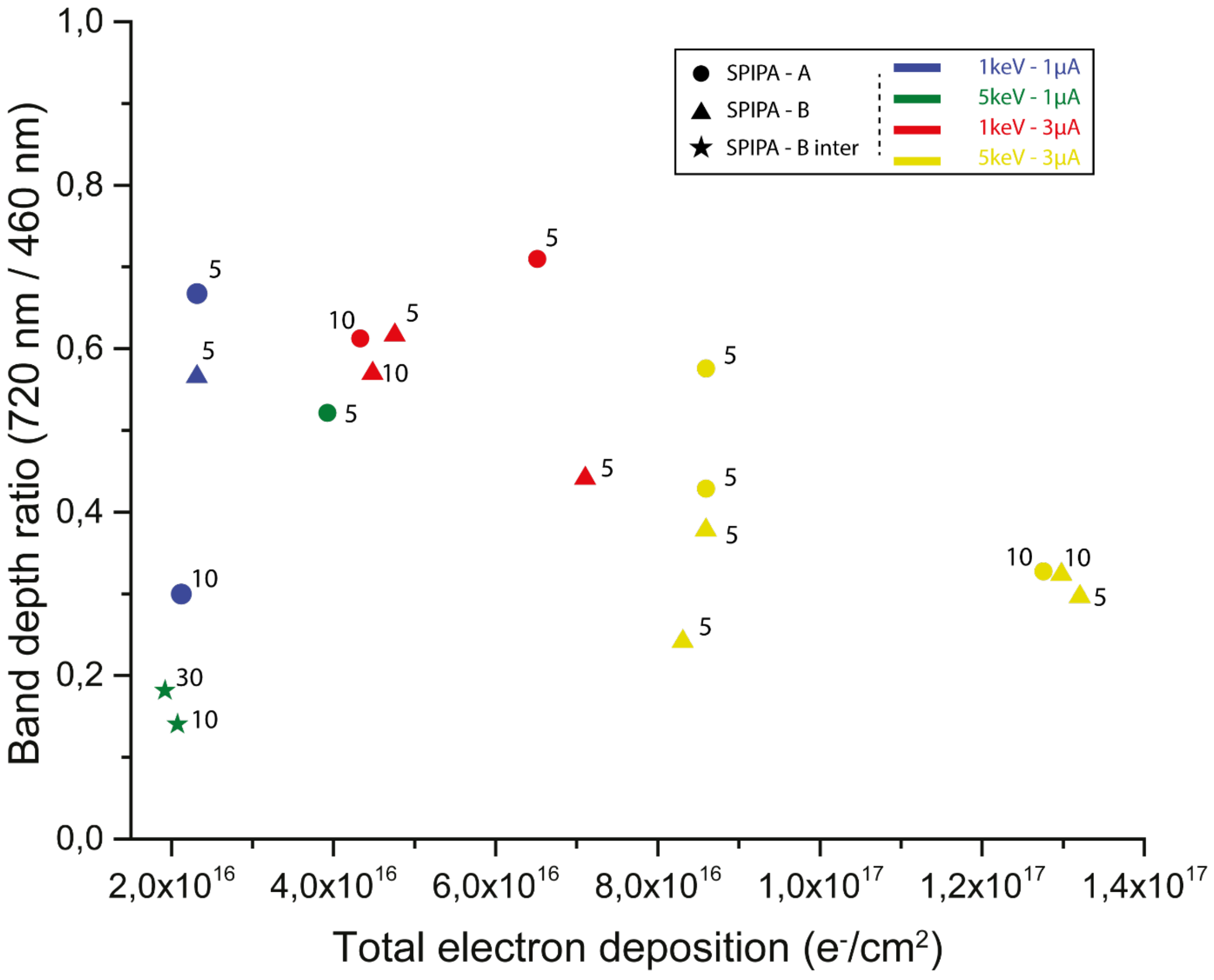

*Figure 13: Band depths ratios of M- and F-centres. The ratios have been computed on a subset of the entire dataset as the M-centre was not clearly observed in all irradiation experiments. The "star" symbols represent the*

*SPIPA-B intermixture described in section 2.1. the number indicated close to the symbols are referring to the salt concentration within the analogue (wt. %). The slabs are not represented as the M-centre has only been observed under stronger irradiations (Section 3.5), which have not been tested on the other analogues.*

It has been previously estimated that the thickness of water ice affected by electron irradiation followed the expression given by (Hand and Carlson, 2011; Johnson, 1990):

$$d \approx R_0 E^{\propto}$$

With E, the energy in keV; $R_0$ = 46 nm and α = 1.76 (empirical determination). It leads to the estimation that a 1 keV beam would affect the top 46 nm, 5 keV, the top 780 nm, and 10 keV the top 2.6 micrometres for pure water ice. (Nordheim et al., 2018) pointed out the potential effect of radiation from centimetres to tens of centimetres below the surface in the most irradiated locations on Europa, but considering the entire range of energies within the Jovian plasma (reaching MeV energies).

### 4.5.2. Implications for Europa's Surface

The number of available visible spectral reflectance datasets for Europa is limited. (Geissler et al., 1998) provide radiance factor values for the six filters (413, 560, 665, 757, 888, and 991 nm) from the Solid State Imaging system on board Galileo (Belton et al., 1992), which was looking at specific features of Europa's surface. Geissler's study reports on the analyses of the reflectance of lineaments (fractures, ridges, double/triple bands) and of the surrounding plains. Their spectra of the "dark" features show the following general tendency: a red slope over the entire visible, the darkest feature being the red-brown unit in the middle of aged double ridges (referred to as triple margins in the paper).

The near-Infrared Mapping Spectrometer (NIMS) of Galileo acquired spectral images from 0.7 to 5.2 micrometres (Carlson et al., 1992), preventing any identification of the F-centres or absorption related to sodium colloids. The observation of the M-centres has, so far, never been claimed.

New observations on the leading side of Europa performed with the Hubble space telescope have recently been reported by (Trumbo et al., 2019). They observed disrupted chaos regions and found a 450-nm feature that they attribute to F-centres due to electrons irradiation with higher energies on Europa's leading side.

The exact modes of association and mixing between water ice and the non-icy components at the surface of Europa are not well constrained. The flash-freezing process is a commonly suggested and assumed process at the surface of differentiated icy moons with global oceans covered by icy shells such as Europa and Enceladus showing respectively indirect and direct evidence of plumes (McCord et al., 2002; Nimmo et al., 2007; Postberg et al., 2009; Roth et al., 2014; Sparks et al., 2016). Our SPIPA analogues mimic the likely outcome of this process and would be relevant for localised observations of plumes' ejecta submitted to radiation. The limitation of the growth of the M-centre within our granular analogues could provide an explanation for the absence of any detection of M-centres at Europa's surface.

Our compact slabs samples with hydrated salts could as well explain the absence of M-centre at the surface of Europa due to the difficulty to produce it in such analogues, as discussed in section 4.2. It is also in good agreement with thermal models of the surface (Rathbun et al., 2010). Producing the one and only F-centre, salty slabs appear to be our most relevant analogue considering Europa's surface VIS spectra available today. However, future high-resolution observations by JUICE and Clipper might reveal regions of the surface where fresh deposits of rapidly frozen salty ice particles share more resemblances with our granular analogues. If such particulate textures are indeed more efficient at producing the features at longer wavelengths (M-centre and Na colloids), the irradiation-induced colouration process will provide interesting opportunities to better identify surface textures –and/or the level of hydration of the salt - at the grain scale from their visible reflectance spectrum.

# 5. Conclusion

We complement previous electron irradiation studies of pure sodium chloride and pure water ice with investigations of analogue samples combining the two components in different ways, mimicking possible surface material at the surface of Europa. Varying different parameters of both the experiments (energy and flux of electrons) and the samples (composition, texture), some of our results resemble the ones of previous campaigns while others show different behaviours. The most important and relevant findings are:

- The counter-intuitive effect of NaCl concentration in both slabs and flash-frozen particles: there is an optimum concentration at which the visual effects are maximised (with the current dataset, positioned around 10 wt%). Further increasing concentration results in a decrease of the absorption features.
- A broad absorption feature related to the formation of Na colloids (580 nm) is exclusively observed with granular analogues. It has also not been reported in previous experiments with pure NaCl grains at cryogenic temperature.
- Pure water ice particles mixed with pure salt particles as inter-mixture react systematically to electron irradiation. It leads to the formation of a strong F-centre absorption, followed by a weaker M-centre absorption, and possibly aggregates of Na colloids. It is an expected result, as the surface of such a sample can be understood as pure NaCl crystals with a contaminant (pure water ice) that is not spectrally affected by irradiation. The irradiation of pure NaCl crystals has been performed in previous studies (Hand and Carlson, 2015; Poston et al., 2017), and these results are in good agreement with their conclusions while the presence of salt crystals as inclusion within a water ice matrix (intra-mixture) induces other behaviour (see the first two points).

- The grain size of the NaCl crystals, hydrated or anhydrous, most likely contributes to the variability of intensity of the absorption features when comparing the F-centre intensities between granular samples and slabs. Comparisons between SPIPA-A and SPIPA-B samples, which differ by their ice particle size (4.5 vs 67 µm diameter) but not by their salt particle size, show that particle size effects are more related to the salt than the ice.

Salty ices containing alkali halides are affected by electron irradiation, as are pure NaCl grains. The resulting effects are nonetheless reduced compared to pure NaCl grains and differ between slabs and granular ices. These results demonstrate the influence of the production method and the resulting variability of texture and composition of the salty ices.

Recent observations of Europa have suggested the presence of the F-centre at the surface, but not the other spectral features observed and discussed in our laboratory work. This could be the result of either (i) a specific type of salty ice structure at the surface (as slab-like surface), (ii) a flux-dependency, or to a less extent (iii) salt concentrations and nature. Finally, a combination of all these parameters in various proportions would be reasonable to assume. Photobleaching has been proposed as a possible explanation to prevent the formation of M-centres, and possibly other crystal defects, by resorbing the F-centres (Georgiou and Pollock, 1989; Mador et al., 1954; Trumbo et al., 2019), and even on a daily basis at the surface of Europa (Denman et al., 2022). This hypothesis is complicated to assess with our dataset as our illumination conditions during hyperspectral cube acquisition were drastically shorter and with monochromatic light.

The irradiation conditions tested in the present study are not covering the entirety of the range of energies encompassed within the Jovian magnetosphere. It is reasonable to assume that the observations conducted in this study may be modified when irradiation is performed with stronger energies, such as the 10 keV available with the MEFISTO chamber. Longer irradiation with lower fluxes of electrons would better mimic the actual irradiation conditions at Europa and would be of major interest to infer the stability of the proposed processes regarding the formation of F-centres and other defects within slabs and granular salty samples. Physical models and simulations will also be necessary to verify some of the hypotheses made here, provide quantitative analyses of the measurements and extrapolate laboratory conditions to Europa surface conditions.

Experimentally, other parameters of the samples could be considered starting with the chemistry. Other species than NaCl, such as $MgSO_4$, $MgCl_2$, and $NaSO_4$ hydrates, are suspected to be present at the surface of Europa and studying the effect of ion and electron irradiation on these compounds in our granular and compact slabs analogues, will be the next step of our laboratory investigations.

## Acknowledgements

The team from the University of Bern is supported by the Swiss National Science Foundation, in particular through the NCCR PlanetS. Bernhard Jost is funded by the NASA Postdoctoral Program Administered by the Universities Space Research Association (USRA). We thank two anonymous reviewers for their careful reading of the manuscript and their useful and constructive reviews.